\documentclass[12pt,preprint]{aastex}

\usepackage{natbib,color}

\newcommand{\angstrom}{\mbox{\normalfont\AA}}

\slugcomment{To be submitted to ApJ}

\shorttitle{Inner Heliospheric Evolution of a ''Stealth'' CME}
\shortauthors{Nieves-Chinchilla et al.}

\begin{document}

\title{Inner Heliospheric Evolution of a ''Stealth'' CME Derived From Multi-view Imaging and Multipoint In--situ observations: I.~Propagation to 1~AU }


\author{T. Nieves-Chinchilla\altaffilmark{1,2}}
\affil{1. Catholic University of America, Washington, DC 20064 \\
2. NASA Goddard Space Flight Center, Greenbelt, MD 20770}
\email{Teresa.Nieves@nasa.gov}

\author{A. Vourlidas}
\affil{Space Science Division, Naval Research Laboratory, Washington, DC 20375}

\author{G. Stenborg}
\affil{George Mason University, College of Science, Fairfax, VA 22030}

\author{N. P. Savani}
\affil{1. University Corporation for Atmospheric Research (UCAR), Boulder, CO, 80307, USA
2. NASA Goddard Space Flight Center, Greenbelt, MD 20770}

\author{A. Koval}
\affil{1. University of Maryland, Baltimore County, MD \\
2. NASA Goddard Space Flight Center, Greenbelt, MD 20770}

\author{A. Szabo}
\affil{NASA Goddard Space Flight Center, Greenbelt, MD 20770}

\and

\author{L. K. Jian}
\affil{1. University of Maryland, College Park, MD \\
2. NASA Goddard Space Flight Center, Greenbelt, MD 20770}




\begin{abstract}
 Coronal mass ejections (CMEs) are the main driver of Space Weather. Therefore, a precise forecasting of their likely geo-effectiveness relies on an accurate tracking of their morphological and kinematical evolution throughout the interplanetary medium. However, single view-point observations require many assumptions to model the development of the features of CMEs, the most common hypotheses were those of radial propagation and self-similar expansion. The use of different view-points shows that at least for some cases, those assumptions are no longer valid. From radial propagation, typical attributes that can now been confirmed to exist are; over-expansion, and/or rotation along the propagation axis. Understanding of the 3D development and evolution of the CME features will help to establish the connection between remote and in-situ observations, and hence help forecast Space Weather. We present an analysis of the morphological and kinematical evolution of a STEREO B-directed CME on 2009 August 25-27. By means of a comprehensive analysis of remote imaging observations provided by SOHO, STEREO and SDO missions, and in-situ measurements recorded by Wind, ACE, and MESSENGER, we prove in this paper that the event exhibits signatures of deflection, which are usually associated to changes in the direction of propagation and/or also with rotation. The interaction with other magnetic obstacles could act as a catalyst of deflection or rotation effects. We propose, also, a method to investigate the change of the CME Tilt from the analysis of height-time direct measurements. If this method is validated in further work, it may have important implications for space weather studies because it will allow infer ICME orientation. 

\end{abstract}


\section{Introduction}
\label{Sec:Introducction}

Coronal Mass Ejections (CMEs) are the largest manifestations of solar transient activity in terms of mass, size, and energy. Although they are not the only cause of geomagnetic activity, they are credited as one of the main drivers of adverse Space Weather. 

Development of reliable Space Weather forecasting requires knowledge of many factors that can affect the evolution of CMEs through the interplanetary medium. To assess their impact to Earth, their interplanetary evolution is usually described under the assumptions of radial propagation and self--similar expansion \citep{Savani2011}. These zeroth-order assumptions are, in general, in good agreement with years of single spacecraft observations of CMEs. However, inaccurate prediction in some cases has serious consequences.

Analyses of white light observations suggest that the change in the CME position angle and angular width \citep[e.g.,][]{Byrne2010} provides clues about the evolution mechanisms. These observations, as projected onto the plane of the sky, are interpreted as CME deflection \citep{Lugaz2012ApJ} or signatures of non--self-similar expansion \citep{Davies2012}. However, in order to shed some light on the true meaning of these observations, de--projection analysis must be performed. Several techniques have been developed to learn about the true kinematic parameters, \citep[e.g.,][]{sheeley1999JGR,Sheeley2008ApJ,Lugaz2009,Lugaz2010SoPh}; or to model the 3D configuration of the CME feature \citep[e.g.,][]{Thernisien2006,Xie2006SpWea,Wood2010ApJ} or to uncover hidden effects such as rotation \citep[e.g.,][]{Vourlidas2011,Nieves-Chinchilla2012}. The use of multi--spacecraft remote sensing observations allows to better constrain various scenarios for the CME evolution, 3D morphology, or determination of the true kinematical parameters. But the debate still goes on. Likewise, multipoint in--situ observations have provided important results for the analysis of the evolution of ICMEs, both radial and temporal \citep[see, e.g.,][]{Bothmer1998,Osherovich1997}.

Lately, with the help of the STEREO mission, the scientific community has embarked on an important effort to link remote sensing with the in--situ observations. Important results from these studies have helped understand complex CMEs scenarios \citep[][]{Mostl2012ApJ} or reveal gaps in the understanding of CME evolution \citep[][]{Nieves-Chinchilla2012,Savani2009}. In all these cases,  the influence of the interplanetary structures interacting with the event under study \citep[see, e.g.,][]{Gopalswamy2001ApJ,Goplaswamy2005GRL,Gopalswamyetal2009JGRA,Savani2010,Wood2012ApJ,Lugaz2012ApJ,Xie2013}, or the pre--existing conditions  \citep{Gopalswamy2004JGR}, seems to play an important role. 

From the statistical work carried out by \citet{Xie2009}, the authors found that slow CMEs tend to deflect towards and propagate along the streamer belts. Also \citet{Makelaetal2013} performed a statistical analysis to a group of Earth--directed ICMEs in order to learn about the influence of the coronal holes on the CME propagation. The ICMEs analyzed were categorized in two subgroups: Magnetic clouds (MC) exhibiting flux--rope signatures \citep[as defined by][]{Burlaga1981}, and non Magnetic Clouds (non--MC) without any flux-rope signatures. The authors found that the influence of Coronal Holes (CH) near the source region (at distances of the order of of 3.2 x 10$^{5}$ km) could deflect an ICME away from the expected direction of propagation. Therefore, their study supports the idea that CMEs can be diverted under the influence of coronal holes. In--situ signatures of interactions among CMEs, CIRs, and ambient solar wind have been identified and studied by several groups, \citep[e.g.,][]{Lepping1997JGR,Burlaga1998JGR,Lugaz2012ApJ}.

Therefore, understanding the 3D evolution of CME features and the influence of the solar and heliospheric environment, is most important to help establish the connection between remote and in--situ observations, and hence improve the forecasting of Space Weather. Here, we undertake a case study of a CME event with many unique characteristics: (i) it exhibits signs of interaction with coronal holes both at the solar source and in the IP space, (2) it is observed by the imagers on STEREO and SOHO and by in--situ instruments at 0.5 AU by MESSENGER and at 1 AU by STEREO, ACE and Wind, (3) it exhibits a very clear flux--rope morphology in the white light images and in the in--situ magnetic field measurements, which facilitates greatly the 3D reconstruction of the event all along its propagation, and (4) it is a typical example of the ''stealth--CME'' class of events \citep{Robbrecht2009}. In summary, it is characterized by extremely weak low--corona signatures, absence of flares, and it is observed to  propagate slowly. 

In this paper,  we demonstrate that the event observed by ACE and Wind along the Sun-Earth line is the same event observed by STEREO-B (STB), $52^\circ$ to the East (Section~2), investigate the kinematical development of the event (Section~3), search for the faint source region(s) (Section~4.1), analyze the latitudinal change observed in the white light images (Section~4.2), and discuss the possible role of the coronal holes in these changes (Section~4.3). We conclude in Section~5.

\section{Event Overview}

\begin{figure}
\includegraphics[scale=0.85]{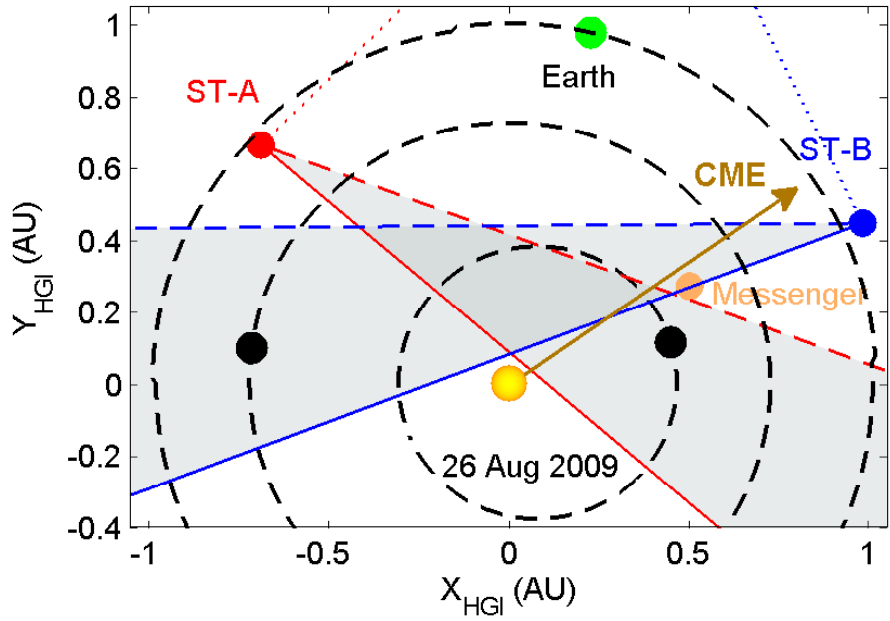}
\caption{Location of STEREO A, STEREO B, MESSENGER, and Earth in the ecliptic plane on 2009 August 26. The STEREO A and B, and MESSENGER spacecraft are at 58$^{\circ}$ (0.96 AU), 52$^{\circ}$ (1.07 AU), and 46$^{\circ}$ (0.56 AU) with respect to the Earth--Sun line, respectively. }
\label{Fig:SPconf}
\end{figure}

The CME and its interplanetary counterpart analyzed in this paper was well observed with several remote--sensing and in--situ instruments on several spacecraft during the time period 2009 August 25--31. In this work, we use the Sun--Earth Connection Coronal and Heliospheric Investigation \citep[SECCHI,][]{Howard2008}  imagers (EUVI, COR1, and COR2) on board the twin STEREO  \citep{Kaiser2008} spacecraft (hereafter STA and STB), and the LASCO C2 and C3 coronagraphs \citep{Brueckner1995} at the L1 point. Magnetic field instruments on board MESSENGER \citep{Solomon2001}, at the time almost in conjunction with STB, also recorded the passage of the interplanetary counterpart of the  CME (hereafter ICME). Figure~\ref{Fig:SPconf} shows the localization of STA, STB, and MESSENGER as of 2009 August 25. During the period of interest, the STEREO spacecraft were separated by $\sim 111^\circ$, STA being $\sim 59^\circ$ ahead of Earth and MESSENGER $\sim 46^\circ$ behind. 
 
The CDAW catalog (http:$//$cdaw.gsfc.nasa.gov/CME$\_$list/index.html) reports a couple of CME events on 2009 August 25. A very narrow CME early on the day, and the one of interest for our work, which is reported to be first seen in the LASCO--C2 field of view (FOV) at 06:30~UT, centered at $82^\circ$ position angle (PA)  with an angular width of $67^\circ$. According to the catalog, it develops in the LASCO--C2/C3 FOV with an average speed of 237~km/s, although the corresponding Height--Time (HT) plot of the outermost part of the leading edge (LE) exhibits a clear acceleration pattern (CDAW reports an acceleration of 8.06~m/s$^2$ at PA $76^\circ$). However, a careful inspection of the LASCO--C2 sequence of images taken on 2009 August 25 reveals that the event starts developing very early in the day, exhibiting a faint brightening and expansion of the streamer on the north--eastern quadrant that leads, in the end, to the streamer blowout late in the afternoon.

The Extreme Ultraviolet Imaging Telescope \citep[EIT][]{Boudine1995}, on board SOHO did not observe any significant activity on the solar disk, other than a small brightening (and dimming of the surrounding area) of a quiet sun region located nearby. Correspondingly, the GOES X--ray channels did not exhibit any increase during the time associated to the origin of the event (i.e., the X--ray flux remained steady at the A--level during the whole day). 

In summary, the event develops very slowly with very weak (and hence hard to detect) on--disk EUV activity. These are typical characteristics of ''stealth CMEs'' \citep[][]{Robbrecht2009}. Because they originate high in the corona, these events do not exhibit low corona signatures at EUV or X--ray wavelengths. The event development between 3 and 15 $R{_\sun}$ is illustrated in Figure \ref{Fig:CorComposition}, where a selected sequence of COR2--A and --B frames is shown. The event appears to propagate towards STB. 

Between 2009 August 25--31, MESSENGER and STB were nearly aligned with the Sun. MESSENGER was located 0.56 AU from the Sun, 46$^{\circ}$ behind Earth, and STB was located at 1.08 AU, 52$^{\circ}$ behind Earth. Therefore, the first to observe the typical signatures of the flux--rope embedded in the ICME was the magnetometer onboard MESSENGER on 2009 August 27. Magnetic and plasma instruments on STB recorded signatures of the passage of the ICME more than 5 days later (see Table \ref{Table:In-situ} and Figure~\ref{Fig:In-situ}). As seen from STB, the event shows the typical signatures of an ICME with a shock in the front driven by a flux--rope. The magnetic structure shows also signatures of plasma compression in the rear part with significant increase of the magnetic field intensity and proton plasma parameters. This kind of effect has been reported before \citep[e.g.,][]{Lepping1997JGR,Burlaga1998JGR} and has been attributed to the interaction of a fast stream compressing the back of the event. Just two hours earlier, ACE observes the passage of an ICME exhibiting similar flux--rope--like signatures and plasma parameters. A detailed description of the in--situ observations is presented in Section~\ref{ICME}.

\subsection{Imaging Observations of the CME}

\begin{figure}
\includegraphics{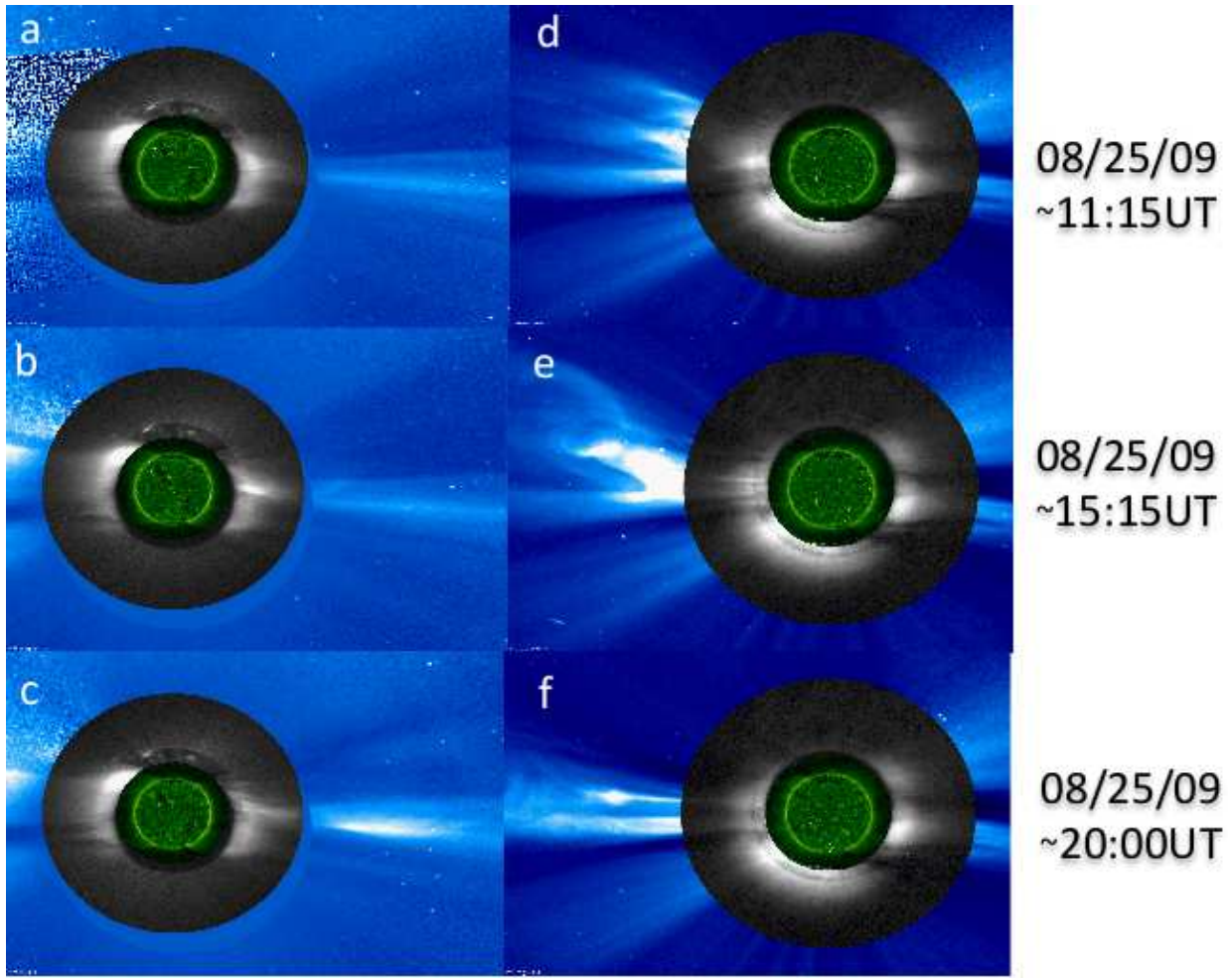}
\caption{Time evolution of the event as observed by STEREO B (left) and STEREO A (right). The composite frames contain images from  EUVI 195\AA\ (green), COR~1 (gray), and COR~2 (blue). (video available)}
\label{Fig:CorComposition}
\end{figure}

The signatures of the event in the COR1--A FOV, and the weakness of the emission in COR1--B indicate that the event is directed, to a large degree toward STB. Therefore, we first looked for signatures of the early stages of the event in STB observations. A slow--rising flux--rope--like structure is seen developing in the EUVI--B FOV above the West limb shortly after 05:00 UT, moving westward. It first appears in COR1--B at 08:00 UT and COR2--B at 14:54 UT (see Figure \ref{Fig:CorComposition}). The images in the EUVI 195 $\angstrom$ channel show the development of an elongated structure above the west limb at around 04:00 UT that matches the temporal and spatial evolution of the feature observed in COR 1B, although nothing noticeable on the disk. At first glance, the EUVI B observations at 195$\angstrom$(cadence: 5 min) and 304$\angstrom$(cadence: 10 min) do not show any signature prior to the observation of the event in the coronagraph FOV that could be associated to the corresponding dynamic feature This fact supports the idea of the stealth nature of the event. However, there is a post--eruption signature (that starts around midday), which is characterized by the heating of the plasma along a neutral line that seems to run along the southern leg of the northern coronal hole. These post--eruption loops are of importance because they allows us to put constrains on the likely source region of the event (see Section \ref{Sec:SolarSource}).  

Interestingly, COR 1A observes the slow development of a flux--rope--like structure above the E limb (slightly North of the equatorial plane) displaying a complex front (Figure \ref{Fig:CorComposition}). The feature is first clearly seen between 03:00~UT and 05:00~UT (due to the kinematical characteristics of the event it is rather difficult to pinpoint the exact time of first appearance). The feature is then first seen in COR~2A at 10:24~UT developing eastwards, without apparently crossing the equatorial plane. Observations of EUVI~A in the 195 $\angstrom$ channel, show the apparent break up of a streamer top between 06:50~UT and 10:20~UT. Unfortunately, it is impossible to narrow down the time of occurrence due to a data gap between those times. In spite of the time uncertainty in the development of the event as observed by EUVI~B in the 171 $\angstrom$ channel, there exist a temporal match between the development of the out--of--limb feature observed in this channel above the W limb and the corresponding event on EUVI-A 195 $\angstrom$. The same applies to COR~1A and 1B observations.  

The CME feature, albeit complex, exhibits signatures on COR1 A typical of a flux rope. The flux--rope signatures remains clear throughout its development on the COR2~A FOV, becoming fuzzy on the first HI~1~A images. Signatures of the flux--rope are first  observed by HI~1~A on 2009 August 25 at 16:09~UT and by HI~1~B on 2009 August 26 at 02:09~UT (a diffuse brightening is observed preceding the structure of interest). The complex morphology of the event is analyzed and interpreted in Sections~\ref{Sec:HT}. 

Several on--disk features were likely to have played a role in the CME development. They will be discussed in detail in Section \ref{Sec:SolarSource}. A detailed analysis of the morphological and kinematical properties of the event is carried out in Sections \ref{Analysis}.

\subsection{In--situ observations of the CME}
\label{ICME}

In--situ instruments onboard STB, in particular IMPACT \citep{Luhmannetal2008_SSRv}, recorded the passage of an ICME on 2009 August 30 exhibiting a typical three--part structure, namely a shock, sheath, and magnetic cloud (MC). Concurrently, a small ICME embedded in a SIR was detected at Earth  \citep[]{Kilpua2011, Jian2011} by the magnetic and plasma instruments on board the ACE spacecraft \citep{Garradetal1998SSRv}. 

\begin{figure}
\includegraphics[scale=0.62]{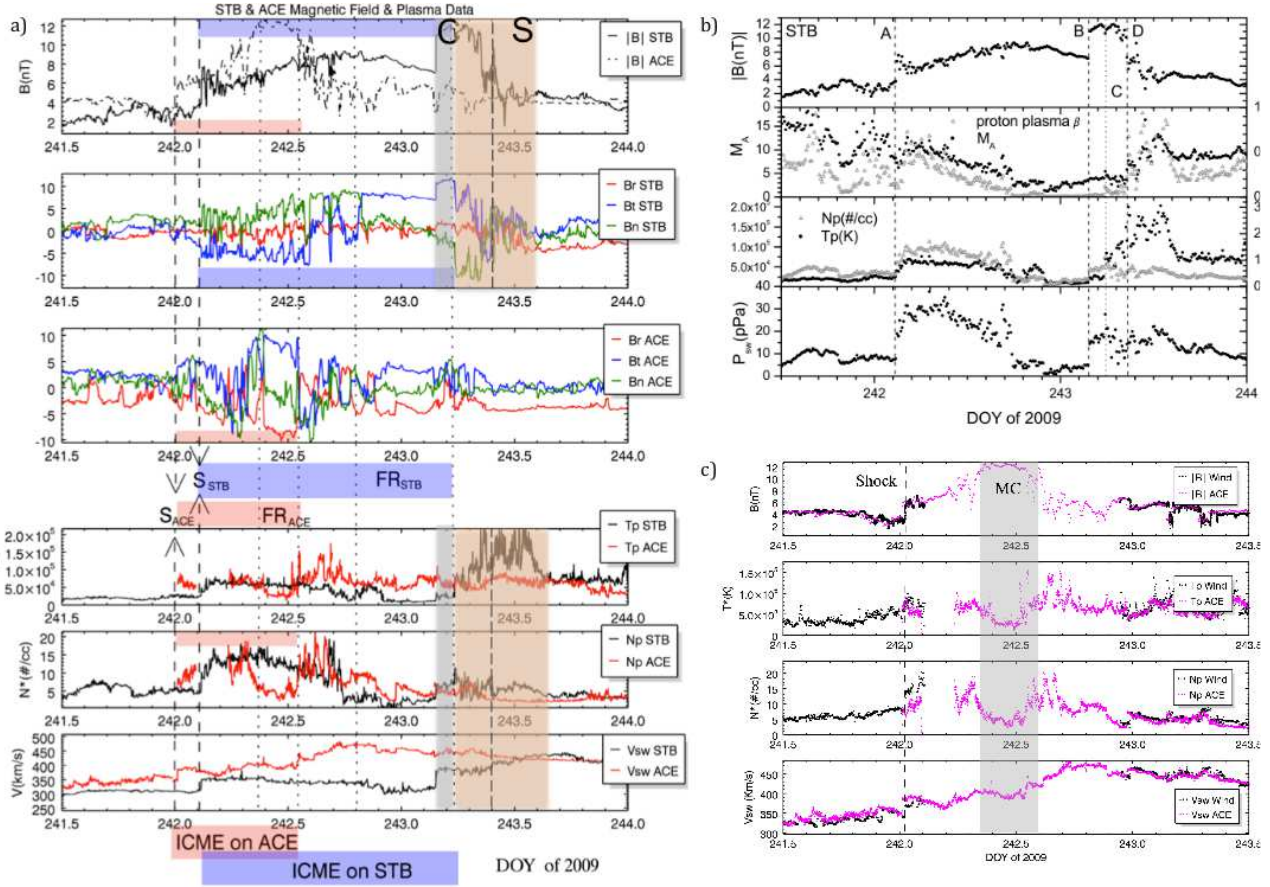}
\caption{Left panel (a): From top to bottom, i) magnetic field magnitude as observed by the magnetometers onboard ACE (dashed black line) and STB (black line); ii) magnetic field RTN components as observed by STB; iii) magnetic field RTN components as observed by ACE. The following three layers are: the solar wind temperature, density, and bulk solar wind velocity, respectively, as measured by ACE (red) and STB (black). The vertical dashed lines indicate the boundaries of the ICME: the interval marked with pink corresponds to ACE, and the one marked in purple to STB. Top right panel (b):  Derived physical quantities for the in-situ plasma and magnetic field parameters as observed during 2009 August 30-31 time interval by STB. Bottom right panel (c): Magnetic field and solar wind plasma parameter from ACE and Wind spacecraft (both located on L1 orbit).  }
\label{Fig:In-situ}
\end{figure}

Figure~\ref{Fig:In-situ}a show the magnetic field measurements and solar wind plasma parameters from 2009 August 29 through 31 as observed by STB and ACE instruments. The magnetic field magnitude is shown in the uppermost panel. The second panel shows the magnetic field components (in the RTN coordinate system) as measured by the STB MAG instrument \citep{Acunaetal2008SSRv}. The corresponding magnetic field components (RTN) as measured by the ACE--MAG instrument \citep{Smithetal1998SSRv} are shown in the third panel. The next three panels include the plasma parameters, namely the proton plasma temperature, the proton density, and the solar wind bulk velocity, respectively, as measured by ACE-SWEPAM (red lines) \citep{McComasetal1998SSRv} and STB PLASTIC (black lines) \citep{Galvin2008SSRv}. The vertical dashed lines delimit the ICME as observed by the different spacecraft, the purple color indicating the passage through STB, and the pink color pointing out the passage through ACE. The shock and flux--rope are labeled between the magnetic field and plasma plots. 

As recorded by STB in--situ instruments, a shock impacts the spacecraft on August 30 at 02:50~UT (Table \ref{Table:In-situ}), followed by a flux--rope like structure starting 16:20~UT exhibiting a North-South rotation of the Normal magnetic field component. A wide sheath ($\sim$13.5 hrs) separating an interplanetary shock from the flux--rope, and another shock inside the flux--rope can also be identified in this ICME. The strong compression in the plasma and magnetic field parameters produced by the second shock is because the ICME was being overtaking by a faster solar wind stream. The ICME--solar wind stream interaction develops a shock (marked as B on Figure \ref{Fig:In-situ}b) at 03:41~UT on August 31 (doy 243). The significant change in the magnetic field magnitude (from 7~nT to 12~nT) is accompanied by a moderate change in the plasma parameters.  The proton plasma temperature shows an increase at 05:40 UT (marked as C on Figure \ref{Fig:In-situ}b). 

There is a data gap in the proton plasma parameters from ACE/SWEPAM until the IP shock onset, which is covered by the Wind/SWE proton plasma observations \citep{Ogilvie1995}, Figure \ref{Fig:In-situ}c. The IP shock on August 30 by 00:16~UT indicates the start time of the SIR, which is supported by the magnetic field data. The mean bulk velocity recorded by STB instruments in the MC interval is 328~km/s with an uncorrected expansion velocity of 16.6~km/s (difference between the front and rear solar wind bulk velocity). On ACE/SWEPAM, the bulk velocity is 401~km/s with an uncorrected expansion velocity of 7.5~km/s. The magnetic field maximum/mean values for both spacecraft are 12.7/ 11.9~nT (ACE), and 12.2/8.6~nT (STB) (see Table \ref{Table:In-situ}). 

\begin{figure}
\includegraphics[scale=.8]{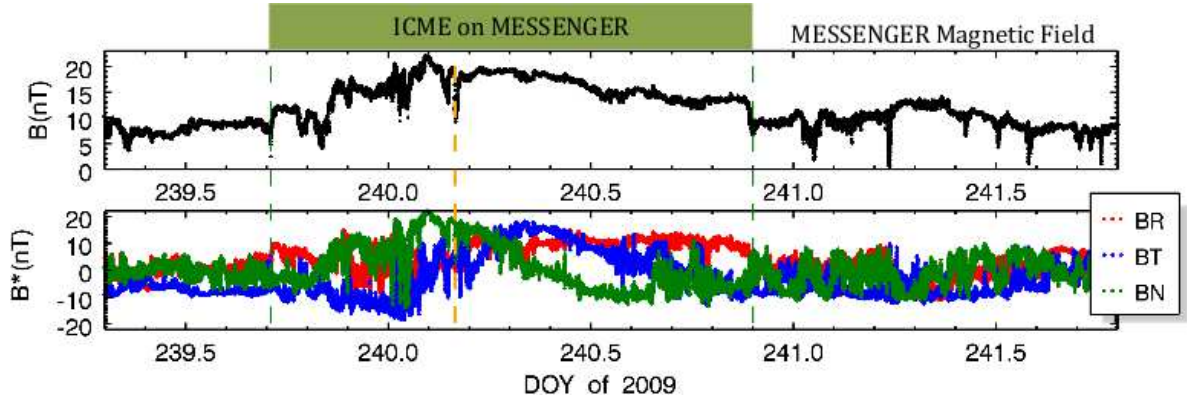}
\caption{MESSENGER magnetometer (MAG) measurements. Top panel: magnitude of the magnetic field (MF). Bottom panel: MF components.}
\label{Fig:Mess}
\end{figure}

As shown in Figure \ref{Fig:SPconf}, MESSENGER was located at 0.56 AU, 46$^{\circ}$ east of the Sun--Earth line, right along the trajectory of the ICME from the Sun towards STB. The magnetic field values (both magnitude and components in the RTN coordinate system) recorded by MESSENGER/MAG \citep{Anderson2007} are shown in Figure~\ref{Fig:Mess}. The plots show an increase of the magnetic field magnitude and a clear rotation of the magnetic field Normal component, the intensity profiles being similar to those recorded by the STB magnetic instrument. The start time of the ICME event as recorded by MESSENGER is signaled on 2009 August 28 at 17:02 UT (Table \ref{Table:In-situ}). The magnetic field is the only signature that marks the ICME boundaries. Table \ref{Table:In-situ} shows the estimated start and end ICME onset time and the boundaries of the embedded flux-rope \citep{Osherovich1997}.  The maximum magnetic field observed was 21.8 nT and the mean value was 15.6 nT during such interval. 

In summary, MESSENGER and STB in--situ data show signatures of a solar transient passing through the spacecraft in agreement with the STEREO imaging observation. The obvious assumption would be that the source region is relatively far from Sun--Earth line and the event will not reach Earth. However, within around the same time interval, ACE in--situ observations at L1 show the passage of a solar transient, headed towards Earth. The timing and the plasma and magnetic field measurements suggest that the STB and ACE transients may be part of a single event. In that case, our original assumption will be wrong. Therefore, to understand the propagation and evolution of this solar transient and verify the idea of a single wide event suggested above, we undertake a comprehensive analysis involving the whole set of observations described above. 

 \section{Analysis and Results} 
 \label{Analysis}

Multispacecraft observations allow us to carry out a comprehensive study of the origin and development of the event. We follow a similar methodology to that employed by \citet{Nieves-Chinchilla2012}. 
In particular, using remote sensing observations from the STEREO/SECCHI suite, we develop in Section~\ref{Sec:HT} a 2D analysis of the mophological and kinematical evolution of the event based on Height--Time (HT) maps of the CME envelope as projected into the plane of the sky of COR~2 and HI~1 instruments. We address then in Section~\ref{Sec:in-situ}, the analysis of the in--situ observations from MESSENGER, STB, ACE, and Wind. We use the results to model a 3D reconstruction of the ICME at different times and solar distances.  

\subsection{2D Analysis: Height--Time  Measurements}
\label{Sec:HT}

Direct measurements of the heliocentric distances of key features on white light images have been the most common method of extracting kinematic information from solar transients as projected onto the plane of sky (POS). This technique allows  to characterize the evolution of the events from their 2D kinematic profiles. This technique, applied to simultaneous observation from distinct vantage points and assisted by the right assumptions (given the optical thin nature of the white light corona, projection effects play a significant role and must be treated carefully), helps interpret the 3D morphology and evolution of the events.

\begin{figure}
\includegraphics[scale=.90]{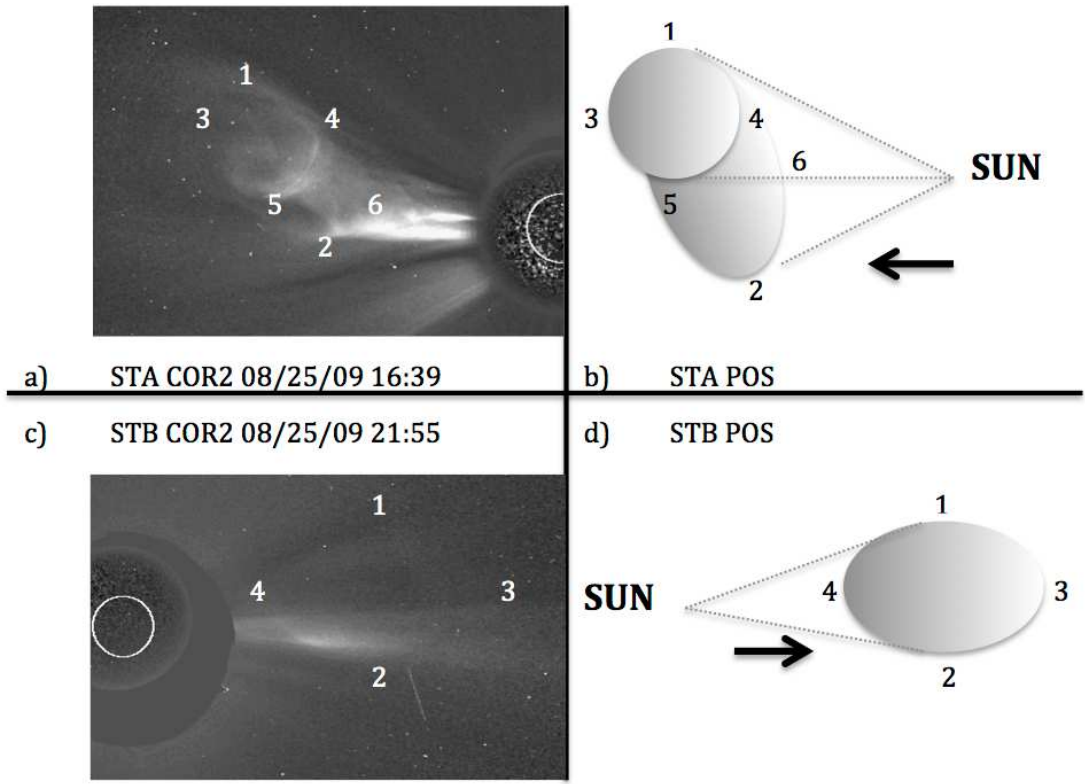}
\caption{Identification of the feature used for the height--time measurements (for details see text). Top Left: COR2-A(16:39~UT on August 25). Bottom left: COR2-B (21:55~UT on August 25). The drawings on the right panels depict a schematic representation of the selected points within the CME, i.e, points 1,3, 4, 5 represent the cross section of the top half, points 1-2 represent the projected extent of the event, etc. Only measurements of points 1,2, and 5 are discussed in this paper.}
\label{Fig:HT1}
\end{figure}

Because of the relative separation of STA and STB, the CME exhibits very different signatures on each coronagraph (see, Figure~\ref{Fig:HT1}, panels a and c). Namely, i) the CME shows up in COR 2A images at higher altitudes than their counterpart in COR 2B; ii) the CME feature as seen by the imagers on STA exhibits a complex structure that hereafter will be simply referred to as flux--rope; and iii) this complex structure does not appear on the imagers onboard STB.

The selected key features tracked in the COR2--A images are shown in Figure \ref{Fig:HT1}a. A schematic cartoon depicting such points is shown in Figure \ref{Fig:HT1}b. The locations numbered [1,3,4,5] in panels a and b were chosen to delimit the projection of what we interpret to be the cross--section of the flux--rope. Jointly with locations [6,2], they delineate the projection onto the plane of sky of the whole CME feature. The projected cross--section of the flux--rope could be followed all across the COR~2A FOV and on a few HI~1A frames. At 02:09~UT on August 26, the signature representing the projected cross--section overlaps with other structures in the HI~1A FOV, making it impossible to track it reliably further. However, the key points delimiting the whole CME could be well identified and therefore tracked until ~07:00~UT on August 26. On the other hand, as stated above, the CME as observed on COR~2B, exhibits a simpler structure. The projection of the CME onto the plane of sky of STB is delineated by the key locations numbered [1,2,3,4] in panels c and d of Figure \ref{Fig:HT1}. 

The tracking of the selected key points allow us to obtain the temporal (radial) evolution of the projected shape of the event. The heliocentric (radial) distance is computed from the elongation at different times of the selected key point assuming a flux--rope CME shaped as a cylinder. The heliocentric distance $x$ is computed assuming the non--curved CME front:
\begin{equation}
x=d_{ST}  \tan \epsilon
\end{equation}
where $d_{ST}$ is the heliocentric distance of the corresponding spacecraft ($d_{STA}=0.96$~AU, $d_{STB}=1.08$~AU), and ${\epsilon}$ is the elongation measured on the respective images. 

\begin{figure}
\includegraphics[scale=.7]{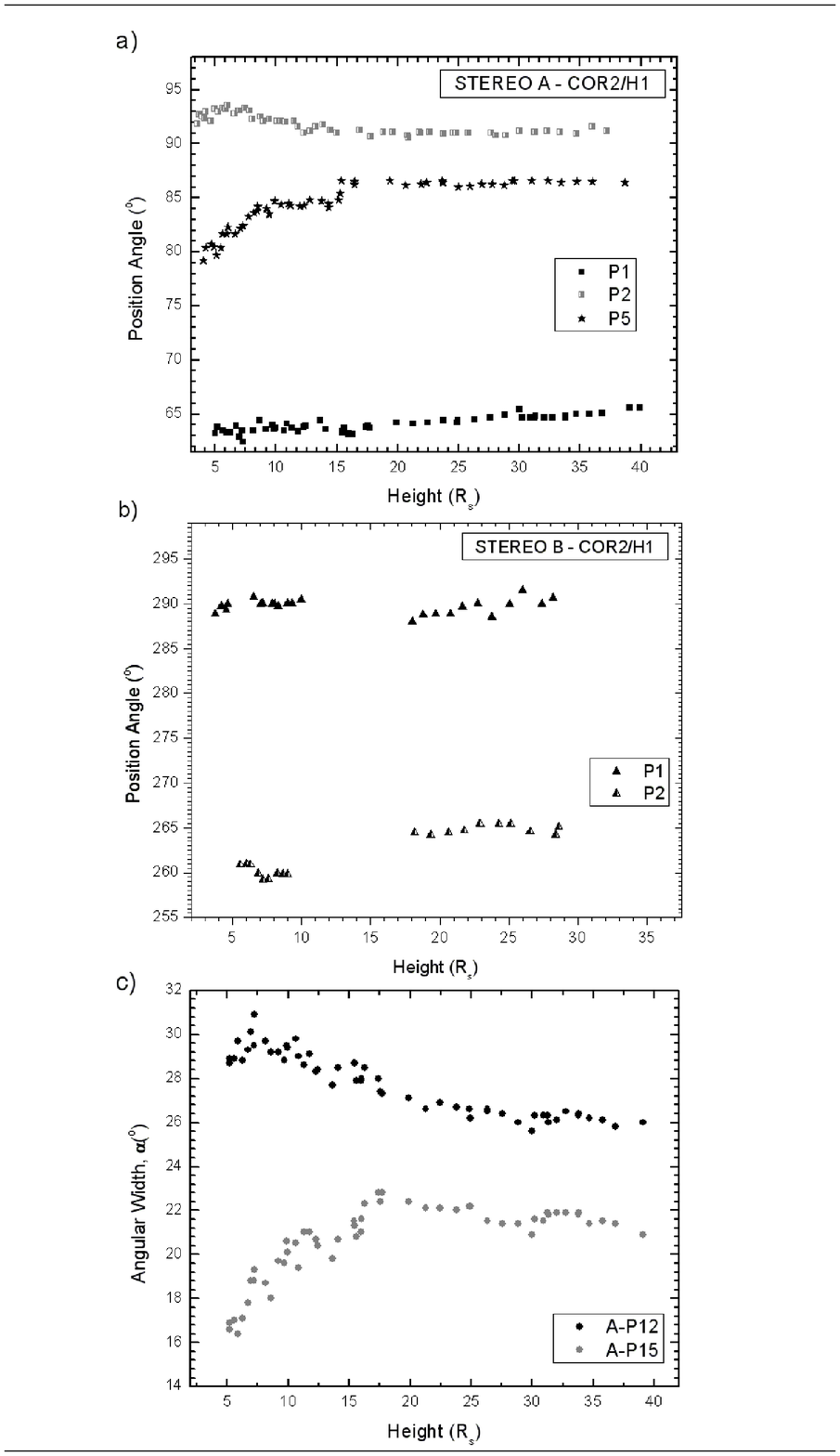}
\caption{a) Position Angle as a function of heliospheric distance of points 1,2,3 on COR2-A and HI1-A (Figure~\ref{Fig:HT1}). b) Points 1 and 2 for the images from COR2-B and HI1-B.  c) Variation of the CME angular width $\alpha$ (ptA21) and cross section of the CME (ptA51) as observed in STA.}
\label{Fig:HT2}
\end{figure}

Figure~\ref{Fig:HT2}, panels a and b, show the corresponding scatter plots of the measurements of the position angle (PA) performed on COR~2 and HI~1 images of the selected key locations: P1, P2, P5 for STA (Figure \ref{Fig:HT2}a); and, P1 and P2 for STB (Figure \ref{Fig:HT2}b). The error of each measurement (not shown in the figure) are different for each instrument. Namely, COR~2 is 2~arc~min, and, HI~1 is 3.6~arc~min. The radial evolution of a derived quantity, i.e., the angular width (defined as the difference in PA between opposite points in the quasi--perpendicular orientation to the direction of propagation) is shown in panel c. The structures identified on Figure \ref{Fig:HT1} are identified on Figure \ref{Fig:HT2} as 'Point' (P) plus the number; for instance, the key point [1] will be the point series (P1) on Figure \ref{Fig:HT2}.  

The small gap observed in Figure~\ref{Fig:HT2}a (i.e., STA) at around $\sim$ 17~R${_\sun}$ corresponds to the end/beginning of the COR 2/HI~1 FOV. Same situation, but larger gap, for STB in Figure~\ref{Fig:HT2}. This is mainly due to the difficulty arisen in following the exact same key point from one instrument to the next due to their diffuseness, and superposition of different structures along the line of sight. Disregarding such small discontinuities on STA, P2 and P5 show a clear increasing or decreasing PA in the COR~2 FOV, while P1 keeps a constant tendency. On H1~FOV, P1, P2, and P5 keep a constant PA (at least within the uncertainty inherent to the measurement). 

\begin{figure}
\includegraphics[scale=0.7]{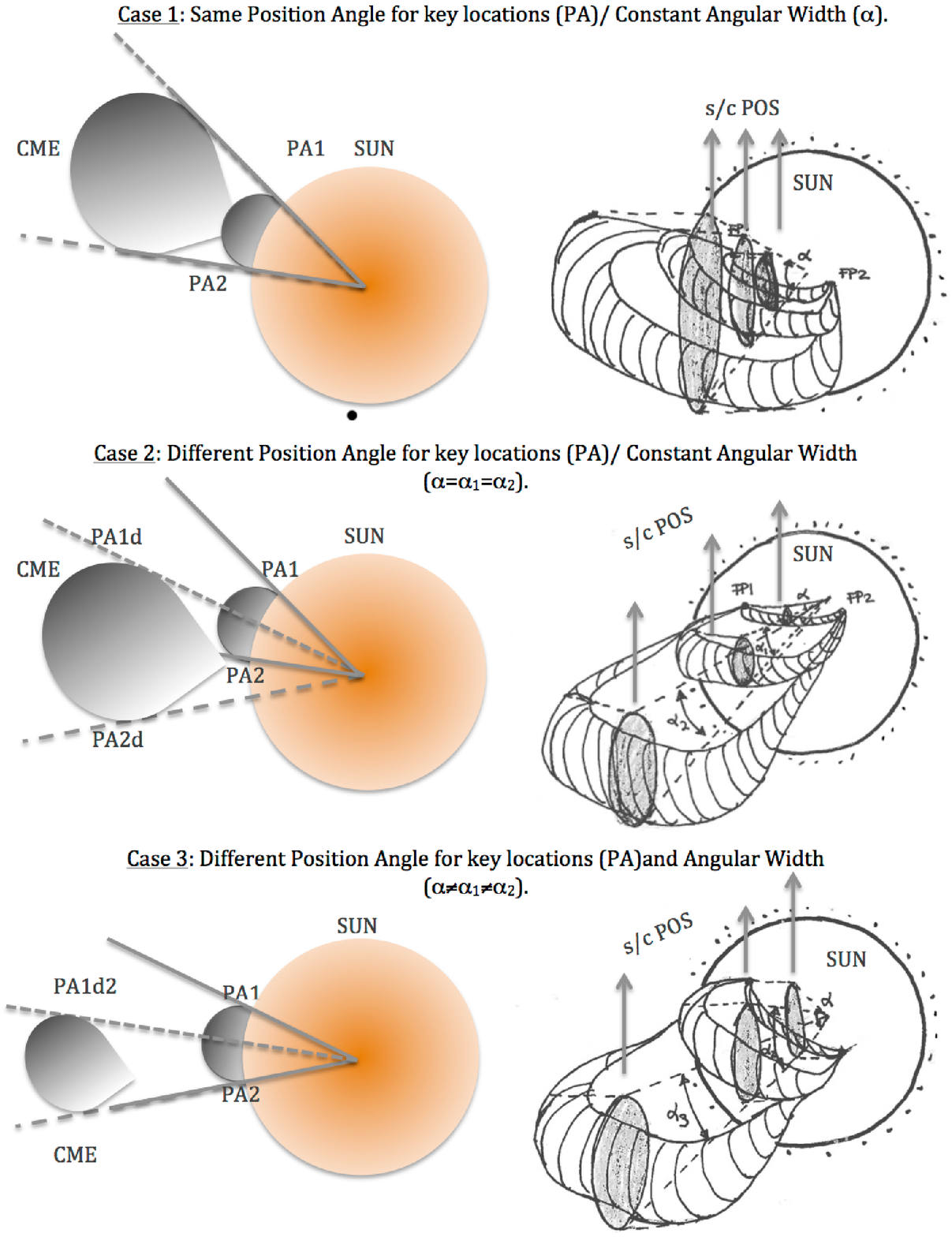}
\caption{Three scenarios that can lead to changes in the CME angular width. Case 1: the top and bottom position angle (PA1, PA2) are constant with the elongation. Case 2: The position angle (PA) of the top (PA1, PA1d) and the bottom (PA2, PA2d) are deflected with the same rate of change. Case 3: The top position angle (PA1) is deflected with the elongation (PA1d).}
\label{Fig:CartoonDef}
\end{figure}

\begin{figure}
\includegraphics[scale=0.8]{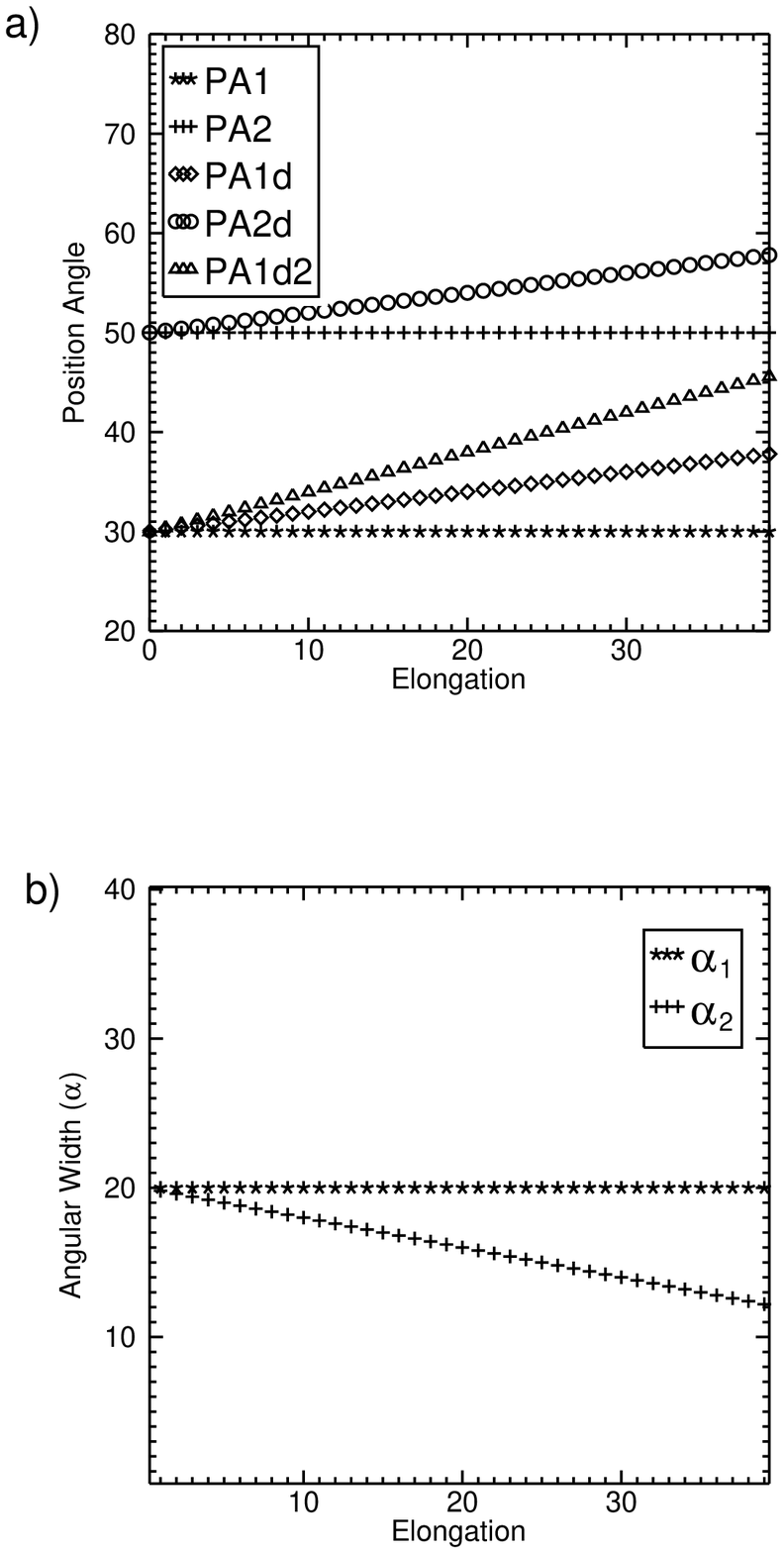}
\caption{a) Position angle (PA) for the top and the bottom for the cases 1, 2, 3 in the Figure \ref{Fig:CartoonDef}. b) Angular width ($\alpha$) for every case in the Figure \ref{Fig:CartoonDef}.}
\label{Fig:DefExplanation}
\end{figure}

We concentrate here our efforts on the interpretation of the change of apparent angular width ($\alpha$) from only STA. To help on this, we summarize in Figure~\ref{Fig:CartoonDef} three possible scenarios of CME development as observed in the FOV of white light imagers; and in Figure \ref{Fig:DefExplanation} the expected dependence of the PA with elongation (top panel) and corresponding derived angular width (bottom panel) for selected key points on the event morphologies depicted in Figure~\ref{Fig:CartoonDef}. For each case depicted in Figure~\ref{Fig:CartoonDef}, there are two drawings. On the left side we show the projected 2D representation, where the PA is represented at two different times for each key location [1] and [2]: initial (PA1, PA2), and end (PA1d, and PA2d). And in the right side, the cartoon delineates the 3D interpretation considering the POS of the S/C located through the CME. A brief description of each case is given as follows: 

\begin{itemize}
  \item \textit{Case 1: Constant PA and $\alpha$}. This is the the simplest and most commonly assumed scenario (self--similar expansion and radial propagation): the PA is constant for key locations [1] and [2] all along the CME development, and therefore $\alpha_{1}$ keeps constant (this case corresponds to the horizontal lines in Figure~\ref{Fig:DefExplanation}). 
  
  \item \textit{Case 2: Different PA but constant $\alpha$}. In this case, the rate of change in PA for both key locations [1] and [2] is the same, and therefore $\alpha_{1}$ in Figure \ref{Fig:DefExplanation}a remain constant with elongation. This case corresponds to the PA1d (circles) and PA2d (diamonds) lines in the top panel of Figure~\ref{Fig:DefExplanation}a, and the horizontal line (star) in the bottom panel, Figure \ref{Fig:DefExplanation}b). Self-similar expansion is a realistic assumption for this case. However, the event is deflected toward the ecliptic plane, away from radial propagation.
 
  \item \textit{Case 3: Different PA and $\alpha$}. In this case, both locations exhibit a different rate of change in the PA. In the case depicted in the bottom panel of Figure \ref{Fig:CartoonDef}, the PA of key location [2] is constant (PA2, Figure \ref{Fig:DefExplanation}a), while the PA of key location [1] varies from PA1 to PA1d2 (triangles in Figure \ref{Fig:DefExplanation}a). The change in the angle is assumed as non self--similar expansion, although it could also be due to the deviation from the radial propagation, away from the POS and/or toward the equatorial plane.
\end{itemize}

Returning to our event, we note that we focus not only on the top and bottom edges of the CME but also on the locations that define the cross--section of the flux rope. The tracking of the key locations, i.e., P1 (CME top edge), P2 (CME bottom edge), and P5 (bottom edge of the assumed cross--section of the flux--rope) shown in Figure~\ref{Fig:HT2}a reveals their variation across the COR2A field of view. 

The position angle of such key locations exhibits a particular behavior: while the top edge of the CME (P1) and the bottom of the \textbf{assumed cross--section (P5)} deflect toward the equatorial plane, the bottom edge of the CME (P2) tends towards the equatorial plane from the South. The temporal variation of the PA of the selected features practically vanishes in the field of view of HI~1 on STA (i.e., between 15~R$_{\sun}$ and 35~R$_{\sun}$).

The changing PA in COR 2A exhibited by P1, P2, and P5 results in a changing angular width of the corresponding structures, i.e, of the whole CME feature, and of the assumed crossed--section of the flux rope. The overall change in the COR 2 FOV is about $\sim4^\circ$ for the assumed cross--section, and $\sim-5^\circ$ for the whole CME (see Figure~\ref{Fig:HT2}c). Therefore, the scenario depicted in {\it Case 3} is the one that can better explain the measurements described in this Section.

\subsection{In--situ analysis}
\label{Sec:in-situ}

As a consequence of the CME evolution in the interplanetary medium, in--situ observations of the solar wind show significant changes \citep[such as in the magnetic field and plasma parameters, energetic particles content, and/or charge states as defined for an ICME by][]{Jian2006}. 

The IP shock is usually taken as the precursor that marks the start time of the ICME interval. Sometimes it is followed by a magnetic topology characteristic of a flux rope, simultaneously with a drop in the proton plasma temperature. These two signatures define a Magnetic Cloud (MC) \citep{Burlaga1981}. Both, the shock and the magnetic cloud are the most suitable entities to be analyzed with different model/techniques.  In this paper, we concentrate mainly on the orientation and geometrical aspect of both the shock and flux--rope. "The analysis of the shock is based on fitting the Rankine Hugoniot  equations to the magnetic field and plasma measurements across the shock using the \citet{Vinas1986} and \citet{Szabo1994} technique. For the orientation/geometry aspects of the flux--rope we use the model developed by \citet{Hidalgo2002a}.

In the case of the event focus of this paper, MESSENGER and STB were almost aligned with the solar source. The flux--rope magnetic field topology is clear and similar from the viewpoint of both spacecraft. However, there are three different elements which are probably due to the CME evolution as observed by the spacecraft. Namely, 1) the compression region in the rear part of the ICME for STB; 2) the wider sheath in STB is due to the value of shock standoff distance varies approximately linearly as a function of \textbf{heliocentric distance \citep[c.f.][]{Savani2012}}; and, 3) the absence of a shock at MESSENGER, while a clear shock is observed at STB (the lack of plasma parameter data could be critical for the IP shock definition, too). Almost simultaneously to STB ICME onset time, ACE observes signatures of an ICME, i.e., a shock, a sheath, and flux--rope like structure. 

We summarize in Table \ref{Table:orientation} the results of the analysis for each IP shock and flux--rope. The axis orientation of the flux--rope measured at MESSENGER location is: longitude ($\phi_{MES}$)=251$^{\circ}$ and latitude ($\theta_{MES}$)=10$^{\circ}$. At STB, the longitude ($\phi_{STB}$) is 253$^{\circ}$ and the latitude ($\theta_{STB}$) is 3$^{\circ}$. Therefore, while the flux--rope appears as laying on the ecliptic plane at STB, the same flux--rope axis at MESSENGER is 10$^{\circ}$ out of the ecliptic plane. The same situation happens with the shock normal direction, i.e., while the shock appears quasi--parallel at MESSENGER, it shows up as quasi--perpendicular at STB.

The analysis of the IP shock--FR at 1~AU shows consistent results. The longitude $\phi_{Wind}$ for the flux--rope is 309$^{\circ}$ with a latitude $\theta_{Wind}$ of -15$^{\circ}$. The shock normal is almost perpendicular to the FR axis.  

\begin{figure}
\includegraphics[scale=.75]{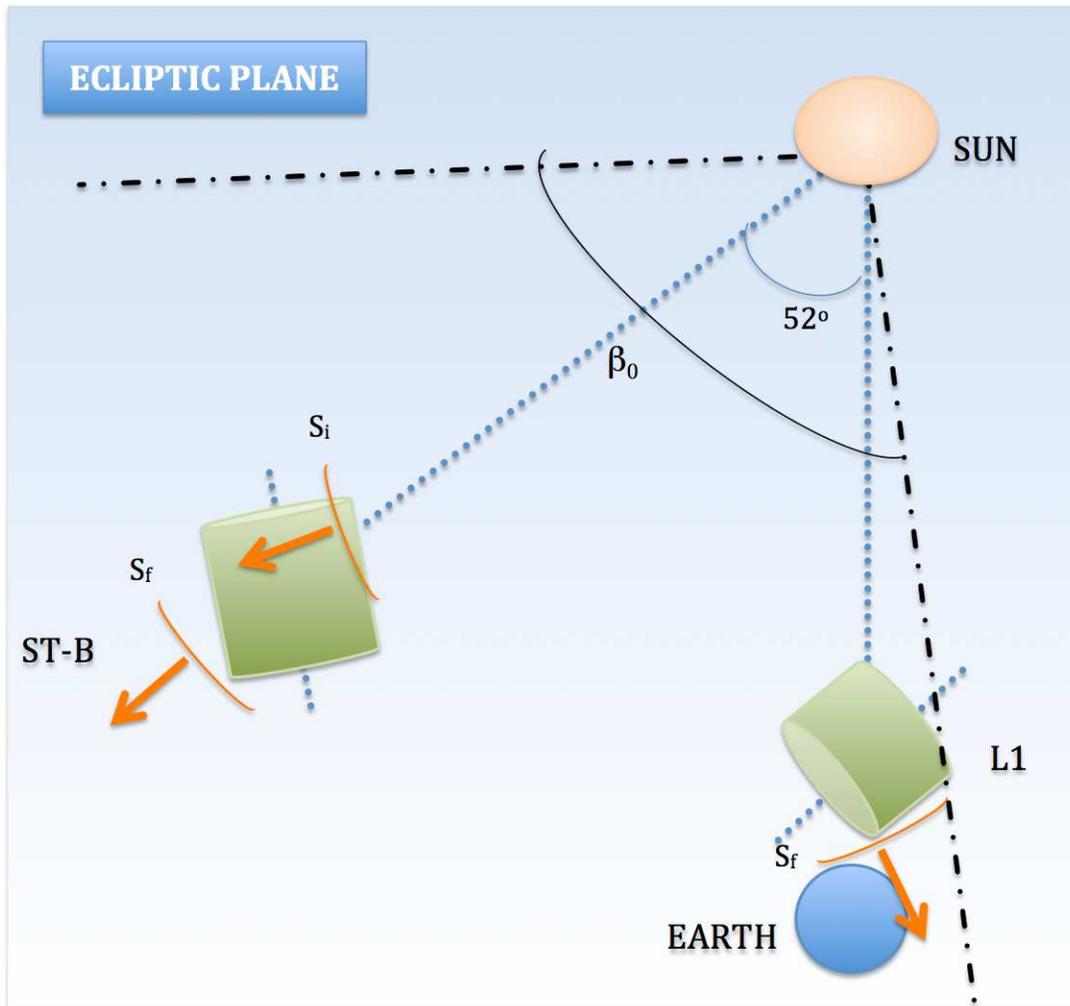}
\caption{Reconstruction on the Ecliptic Plane of the ICME as observed by STEREO B, and ACE-Wind. The flux-ropes are oriented with respect to each spacecraft. The arrows indicate the direction normal to the IP shock. The dash-dotted line marks the extent of the CME source region projected on the Ecliptic.}
\label{Fig:InSituConf}
\end{figure}

The reconstruction on the ecliptic plane of the ICME as observed by the in-situ instruments onboard STB and ACE (Wind) is represented in Figure~\ref{Fig:InSituConf}.

\section{Discussion}

To shed light onto the physical mechanisms at work during the propagation of the CME through the inner heliosphere, we build a timeline of events linking the remote sensing to the in--situ observations.

So far, we have analyzed the remote sensing observations obtained with the white light imagers onboard STEREO and the in--situ observations from MESSENGER, STB and ACE/Wind. The height--time measurements of the event reveal that the CME exhibits an observable change in its angular width, $\alpha$, during its evolution mainly across the COR~2A FOV. This change can be explained by the scenario presented in \textit{Case 3} (Figure \ref{Fig:CartoonDef}, Section~\ref{Sec:HT}). The in--situ reconstructions of the ICMEs observed by STB and ACE/Wind (Section~\ref{Sec:in-situ}) at 1 AU are also consistent with this deflection scenario; namely, the ICME lays on the ecliptic plane and the ICME front seems to be closer to STB than ACE. 

In this Section we will investigate in detail whether the angular width change is a result of the change in the CME \textit{Tilt}, and whether the remote sensing observations are consistent with the in--situ analysis. We will proceed in two stages. First, we search for the likely solar source of the event, in spite of the stealth nature of the event, by looking for post--eruption signatures on the disk. Next, we fit the angular width change and predict the orientation of the CME at its encounter with ACE and STB to connect the imaging observations  with the 3D reconstruction derived from the in--situ analysis. Finally, we interpret the in--situ observations in light of  the CME interaction with the solar wind and nearby coronal holes.

\subsection{Source Region Identification}
\label{Sec:SolarSource}

Stealth CMEs are a distinctive category of CMEs characterized by the lack of surface signatures which makes it difficult to investigate their origins and initial extent or propagation direction. In our experience, however, post--eruption signatures for these events can be found after a careful search. They can provide very useful information about the initial orientation, direction and size of these events as they did in this case. 

Despite the well--defined flux--rope appearance of the CME in the coronagraph images, the identification of its source region was very challenging. Because of the halo appearance in the COR2-B images, we suspected that the source region should lay close to the STB central meridian and hence near the STA sky plane. Therefore, we turned first to the EUVI--A images to search for any evidence of outflow over the STA East limb. As usual in our analyses, we wavelet--processed the EUVI images to enhance the faint off--limb emissions \citep{Stenborg2008}. Indeed, the EUVI--A 171 and 195 $\angstrom$  images show evidence of slowly rising loops in some areas and changing topology in other areas, all along the STA East limb. Using EUVI--COR1-COR2 composite movies (a frame is shown in Fig.~\ref{Fig:CorComposition}) for both STEREO spacecraft, we were able to reconstruct the initial stages of the eruption as follows. 

\begin{figure}
\includegraphics[scale=0.5]{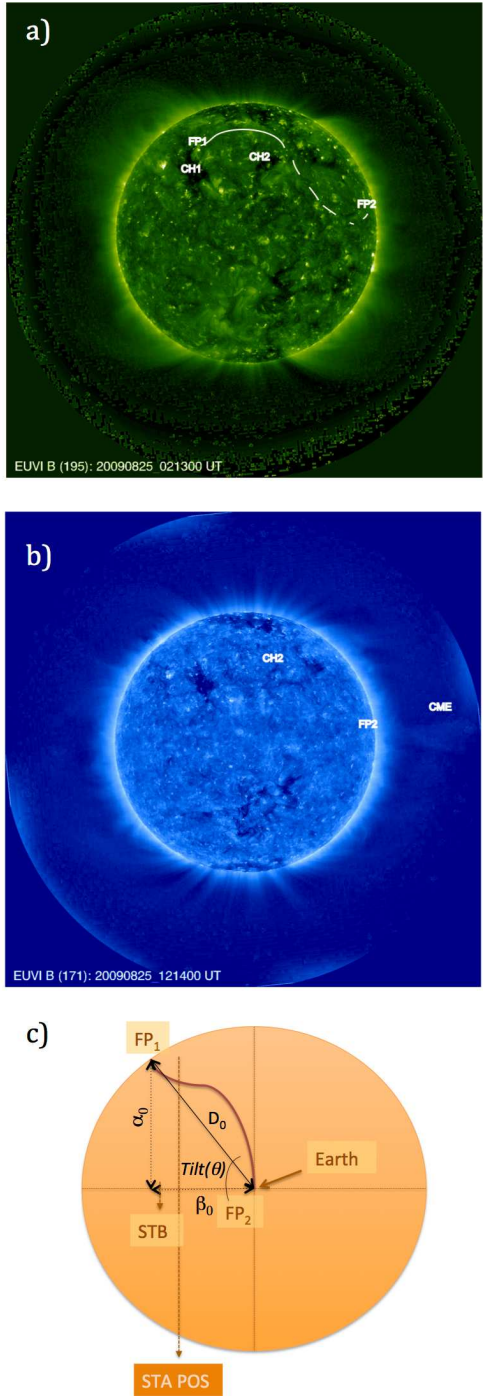}
\caption{a-b) EUVI-B 195 $\angstrom$ image (from online movie) showing the identification of Footpoint 1 (FP1) at 2:14 UT and the extent of post-CME loops (solid line). The dotted line marks our estimate for the total extent/geometry of the event. The other CME footpoint, FP2, was identified at 12:14 $\sim$ UT using the 171$\angstrom$ image below. b) EUVI-B 171$\angstrom$ image (from online movie) at the estimated time of FP2 lift-off at 12:14 UT. The bottom half of the CME is marked with ''CME''. The two coronal holes, CH1 and CH2, are marked in both frames. c) Schematic geometry of the CME source region as seen from Earth. The parameters $\alpha_0$, $D_0$, $\beta_0$, and $\theta$ are the CME initial angular width as seen by COR2-A, extent, extent projected on the Ecliptic, and tilt, respectively. (also available the STA movie)}
\label{Fig:FP}
\end{figure}

In COR2-A (Figure~\ref{Fig:HT1}a), the event presents the morphological characteristics of two flux ropes, one lying close to the ecliptic (\textit{bottom-half}, hereafter) and the other propagating northward (\textit{top-half}, hereafter). \textbf{Similar case was reported by \citet{Deforest2013ApJ}}. This ''duality'' is a result of imaging the two ends of a single, but inclined, flux rope, as we will discuss later. The top half is seen first in the COR1--A images on August 25 at around 3:30~UT at the NE limb. The EUVI--A 195 $\angstrom$ images show that a set of loops, at the approximate position angle of the CME, is blown open between 22:14~UT (August 24) and 02:14~UT (August 25). Based on that detection, we are able to identify a candidate for the source region on the EUVI--B images via the tracing of a small brightening (at 2:14~UT) at Carrington $279^\circ$ longitude and $42^\circ$ latitude (see Table~\ref{Table:SunElem}) right next to a coronal hole (CH1, hereafter). We refer to this as the first footpoint, FP1 (Figure \ref{Fig:FP}a). The extent of the  brightening is too small to account for the size of the observed \textit{top-half} of our event and so it could only represent the extreme eastern end of the ejected structure. The EUVI--B images show faint extended brightening, consistent with post--eruption loops, along the northern coronal hole boundary, moving approximately NW to SW along the polar coronal hole boundary (see trace in Figure~\ref{Fig:FP}a) but this brightening occurs a little later at 12:14~UT. These observations suggest that: (1) the eruption occurred sometime around 0~UT and evolved for about 12 hours, and (2) that the orientation of the source region of the \textit{top-half} lies approximately East--West. This orientation is consistent with the COR2--A view of a circular flux--rope--like structure.

The EUVI--COR1--COR2--A composite images show that the source region of the \textit{bottom-half} is associated with a set of slowly disappearing loops above the east limb of STA at equatorial latitudes. These loops disappear completely by 12:14~UT. The simultaneous EUVI--B 171  $\angstrom$  image (Figure~\ref{Fig:FP}b) shows a faint V-shape feature over the STB West limb (labeled ''CME''), consistent with the backend of a flux rope oriented parallel to the equator. The composite movies verify that this feature is indeed the \textit{bottom-half}. Tracing the footpoint of this feature is not straightforward but there is some activity at the on--disk projection of this feature ($14^\circ$ lon, $3^\circ$ lat). We refer to this as the second CME footpoint, FP2. 

The initiation of the CME is very gradual, lasting for more than 12 hours. The features that could play a role in the CME evolution are marked on Figure \ref{Fig:FP} (panels a and b). The Stonyhurst coordinates for each feature of interest are given in Table~\ref{Table:SunElem} for  both the estimated time of the beginning of the eruption and corresponding ending time. We emphasize that our selections for the footpoints, and hence the extent of the structure, are upper limits. The plane of sky of STA (STA POS), as seen from Earth, and other geometric parameters of interest are indicated in Figure~\ref{Fig:FP}c.   We use FP1 and FP2 to derive a rough estimate of the CME's initial size and orientation: 
\begin{itemize}
\item ~ CME \textit{Tilt} ($\theta$): 22$^{\circ}$; 
\item ~ Projection onto the STA POS (the CME angular width $\alpha_{0}$ as seen from STA): 39$^{\circ}$; 
\item ~ Projection onto the ecliptic plane ($\beta_{0}$): 95$^{\circ}$;
\item ~ Approximate width of the CME source region (D$_{0}$): ~102$^{\circ}$. 
\end{itemize}
\textbf{The value of the CME width projected on the ecliptic plane ($\beta_{0}$) is an important constraint for the in--situ reconstruction}. It is shown in Figure \ref{Fig:InSituConf}, and it is assumed constant during the CME propagation to 1~AU. In that case, the in--situ reconstruction should be constrained inside this cone if radial propagation is assumed. These measurements predict that the Earth will cross the flank of the CME.


The value of $\alpha_0=39^\circ$ as estimated by the source region extension, is slightly greater than the value of $\sim 26^\circ-30^\circ$ derived from the height--time measurements in COR2--A and HI1--A (Figure~\ref{Fig:HT2}c) and gives us confidence in the identification of the source region because it is consistent with the downward trend seen in the white light images. However, the small discrepancy could be due to the uncertain location of FP1 or FP2. 

We have also identified two coronal holes in Figure~\ref{Fig:FP} located on opposite sides of FP1. The coronal hole to the East, CH1, shows little variability during the CME lift--off. However, the coronal hole to the West, CH2, evolves quickly. It reaches a maximum size around 2:00~UT, the estimated time of the FP1 disconnection and vanishes by 04:00~UT on August 26. The coordinates of the coronal holes at the estimated lift--off times of the footpoints are shown in Table~\ref{Table:SunElem}. CH1 reaches the sub--point STB longitude $308^\circ$, at the time of the eruption, on August 27 by 12:00 UT. 

\subsection{Deflection/Rotation}
\label{Sec:deflection}

In general, variations of the CME position angle are atrributed to deflections toward the ecliptic plane and/or away from the radial propagation, i.e., to changes in latitude and/or longitude \citep[][]{Makelaetal2013}. Variations in the CME angular width, on the other hand, are rare. We have argued in the past \citep{Vourlidas2011} that such changes are associated with rotation. This event is no different, although we cannot talk of rotation in this case, because the angular width changes very slightly. It is more  more appropriate to talk about a \textit{tilting} CME. We will show now how to link the change observed in angular width with a change of the CME \textit {tilt}.  

First we note that the angular width of the CME decreases (Figure~\ref{Fig:HT2}c, A-P21) while its assumed CME cross section increases
(Figure \ref{Fig:HT2}c, A-P51), both features exhibiting a similar rate of variation. This behavior can be understood with the help of
the cartoon of \textit{Case 3} in Figure~\ref{Fig:DefExplanation}. As the CME \textit{Tilt} ($\theta$) decreases, the information about the
CME longitudinal extent, which contributes to the P12 measurement, and hence to the projected angular width, is lost. At the same time, the
projection of the CME cross section (P15) on the STA POS is minimized resulting in larger values. In general, the variation of the angular
width $\alpha$ can be represented by a general power law:
\begin{equation}
\alpha=\alpha_{0}r^{d} 
\label{Eq:alpha}
\end{equation}
where $\alpha_{0}$ is the projected angular width and \textit{d} accounts for the rate of change of the angular with heliospheric distance. Assuming a constant CME extent, \textit{$D_{0}$}, the CME Tilt $\theta$ should decrease if $\alpha$ decreases, because the following geometrical relationship applies:
\begin{equation}
D_{0}=\frac{\alpha_{0}}{\sin\theta_{0}}=\frac{\alpha}{\sin\theta}
\label{Eq:theta}
\end{equation}
where $\theta_{0}$ is the CME Tilt based on the source region
orientation on the solar surface. 

\begin{figure}
\includegraphics[scale=1.9]{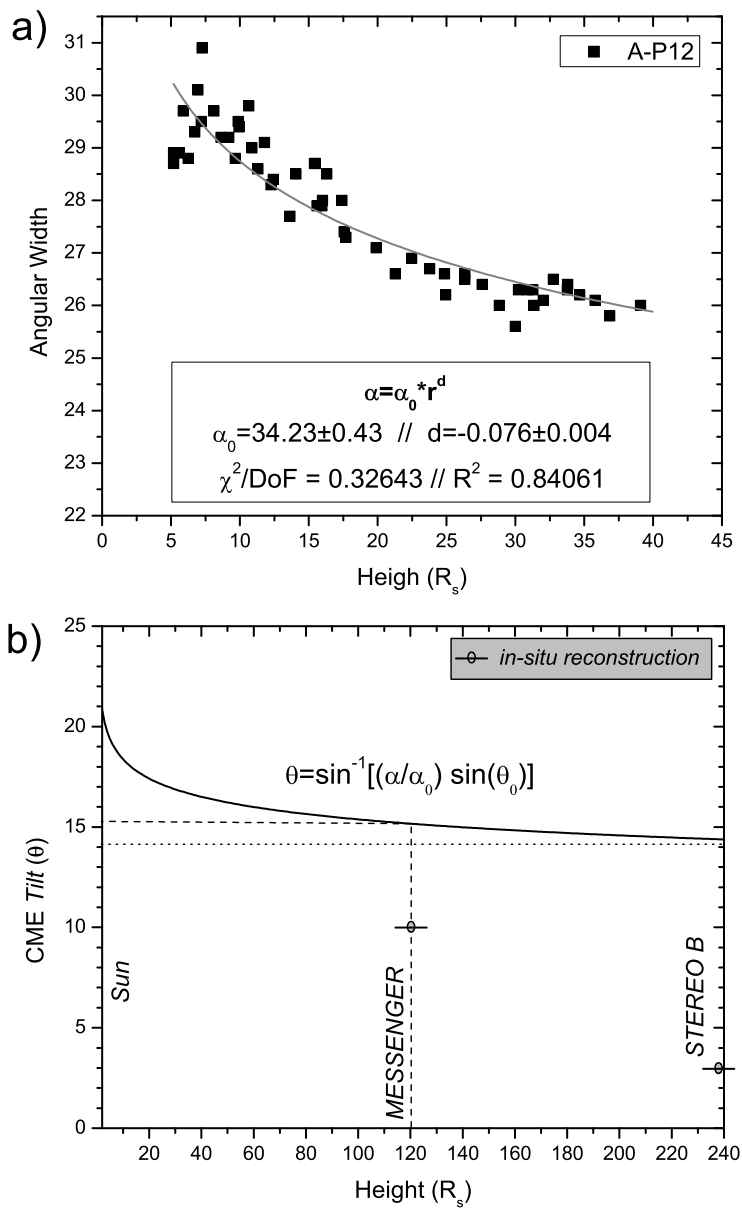}
\caption{a) Power-law fit (grey line) of the CME Angular Width ($\alpha$) measurements (black squares) from COR2-A and HI1-A on August, 25. The fitted parameters and goodness-of-fit are also shown on the panel. b) The predicted CME tilt, $\theta$, change as a function of heliocentric distance. The values from the in-situ reconstructions of the MESSENGER and STB data are shown for comparison. }
\label{Fig:TiltAAA}
\end{figure}

From equations \ref{Eq:alpha} and \ref{Eq:theta}, the value of the CME
tilt as a function of heliospheric distance is
\begin{equation}
\theta=\sin^{-1}[r^{d}\sin\theta_{0}]
\label{Eq:tilt}
\end{equation}
Figure~\ref{Fig:TiltAAA}a shows the result of fitting the CME angular
width, $\alpha$, to equation~\ref{Eq:alpha}. The fit is 
\begin{equation}
\alpha_{0}=34.23^{\circ}\pm 0.43^{\circ}, \quad  d=-0.076\pm 0.004 
\end{equation}
It can be seen that the fitted $\alpha_{0}$ value is very close (to
within $5^\circ$) to the width based on the size of the source region
($39^\circ$, Section~\ref{Sec:SolarSource}). 

Figure \ref{Fig:TiltAAA}b shows the extrapolated CME \textit{Tilt} to 1~AU using Equation~\ref{Eq:tilt}. The predicted \textit{Tilt} at the
MESSENGER position ($\sim$15$^{\circ}$) is again within $5^\circ$ of the tilt obtained by the in--situ reconstruction ($\theta_{MESS}$=10$^{\circ}$). However, the extrapolation of the \textbf{titl} to 1AU ($\sim$14$^{\circ}$) is not as consistent with the in--situ analysis (i.e.,  $\theta_{STB}$=3$^{\circ}$ and $\theta_{Earth}$=-15$^{\circ}$). 

As seen from Figure~\ref{Fig:TiltAAA} (panel b), the model predicts that most of the tilt occurs during the first 60~R$_{\sun}$. The rate
of variation of the CME Tilt derived from the model gradually decreases afterward, and the predicted tilt value tends to the values
obtained by the in--situ models. It seems, therefore, that Equation~\ref{Eq:alpha} and measurements of the CME angular width from
imaging observations close to the Sun offer a practical technique to extrapolate the CME tilt in the inner heliosphere and provide another
means to validate the results of in--situ reconstructions.

\subsection{The Role of Coronal Holes in the CME Tilt Variation}

In the previous section, we were able to estimate the change of the
CME \textit{Tilt} from direct measurements on coronagraph white light
images. However, we have not addressed yet the possible causes for
this change. Based on our measurements and the location of the source
region for the \textit{'top-half'} of the event, it is clear that the
tilt is caused by the equatorward deflection of this CME part. This is
expected behavior for such types of ejections. It
originated in an area flanked by three coronal holes (CH1, CH2, and
the polar CH), and propagated at a very low speed. The polar coronal
hole likely limits the propagation of the CME below $65^\circ$
position angle. CH2 is, however, enveloped by the CME, if our source
region identification is correct in Figure~\ref{Fig:FP}a. CH2 is
short lived and disappears on August 26 at around 04:00~UT. The
timing of the events suggests that between
the CME initiation and its development to 40 R$_{\sun}$, CH2 could
have affected the CME evolution. The variation of the angular width,
$\alpha$, occurs in this time range and could be a result of the
interplay between  the equatforward influence of the polar coronal hole
and the northward influence of CH2, since the CME cannot penetrate
through those open magnetic fields. Such ''tug-of-war'' could indeed
prevent the whole CME from moving toward the streamer belt as usually
seen and cause the rotation signatures we see in the COR2--A images. 

The other coronal hole, CH1, may have played a role later in the CME
evolution. CH1 was long--lived and retained its size from the early
beginning of the CME until at least its detection with the in--situ
instruments at 1 AU. It was located slightly below (southward) of the
CME \textit{top-half} and reached the STB sub--point on August 27 at
around 12~UT. At that time, CH1 was approximately behind the CME
section directed toward STB and was located low enough ($32^\circ$
north) so that the fast solar wind from CH1 could interact with the
backend of the CME at some point. MESSENGER magnetic field
observations (Figure \ref{Fig:Mess}) on August 28, do not show
any signatures of interaction between the fast stream wind and the CME
but this is expected given the short time available for interaction
(about half a day). On the other hand, STB magnetic field and proton
plasma parameters indicate such interaction on August 31, as we have
already discussed in Section~\ref{Sec:in-situ}. Therefore, the contact
between the CME and the fast stream must have happened in the region
between MESSENGER and STB. Indeed, using the calculated speed range
for the fast stream from CH1 (299 - 577)~km/s and the solar wind
velocity observed at STB (425~km/s), we find that the estimated time
of contact between the CME and the fast stream was on August 29
at $\sim$02:00~UT.
  
\section{Summary and Conclusions}

The heliospheric evolution of CMEs, under the simplest approach,  is generally described under the
assumptions of radial propagation and self--similar expansion. Both
assumptions imply constant CME position angle (PA) and/or angular
width ($\alpha$). Thus, any variations of the CME position angle are
attributed to deflection toward the ecliptic plane and/or away from
radial propagation. In this paper, we have investigated a different
possibility for such observations; namely, the variation of CME tilt.

Our event is directed towards STB. The slow
and gradual early development commenced on August 25, and the CME
reached STB at 1~AU on August 31. The event could be tracked remotely
by STEREO, and was detected with in--situ instruments onboard
MESSENGER, STEREO, Wind and ACE. The event was selected because of its
very clear morphology in the coronagraph which allowed us to
measure both its width and its lateral extend. Using the Height--Time
measurement analysis of key features within the event as observed by
the white light imagers on STA, we found a small but coherent and
gradual variation of the event angular width $\alpha$ and cross
section. The two parameters varied in opposite sense with the angular
width reducing and cross section increasing as a function of
heliocentric distance. This unusual behavior can be understood simply as a
change of the CME tilt. 

To prove this scenario, we identified the CME source region and showed
that it implied a very extended eruption ($\sim 95^\circ$),
inclined $22^\circ$ relative to the equator. The extent of the source
region was consistent with the detection of this event as a magnetic
cloud by both ACE and STB, $52^\circ$ to the east. We fit the observed
CME tilt with a power law and computed the predicted tilt at the
location of MESSENGER, STB, and ACE. The predictions were very close
to the in-situ reconstructions results for  MESSENGER ($\Delta\theta=
5^\circ$) and consistent with the 1 AU results ($\Delta\theta=
11^\circ$ for STB). These encouraging
results suggest that it may be possible to predict the CME tilt from
imaging observations close to the Sun and hence provide a useful
constrain to in-situ reconstruction of magnetic clouds and the to
estimate of the orientations of the magnetic field within these
structures. 

To investigate the causes of the tilt, we looked at the possible
influences from two nearby coronal holes, named CH1 and CH2. Both of
the holes were located slightly southward of the erupting flux rope. We
suggested that the short--lived CH2 may have counterbalanced the
equatorward driving of the north part of the CME by the strong polar coronal hole
resulting in the observed tilt. We also suggested that CH1 may have
been responsible for the strong compression seen at the back of the MC
when it was detected by STB. The slightly southward location of CH1
relative to the CME fluxrope may have allowed the rotation of the CH1
under the CME and enabled the subsequent interaction. 

In summary, we have proposed a method to investigate the change of the
CME \textit{Tilt} from the analysis of Height-Time measurements
(Section~\ref{Sec:HT}). We found that the results are
consistent with in-situ reconstructions and solar source
analysis. If this result is validated in further work, it may have
important implications for space weather studies because it will allow us to infer the ICME
orientation at 1~AU using remote sensing observations of the first
stages of the CME. We have proposed a possible explanation for the CME
tilt as interaction of the slowly erupting CME with the opposite
directed fast streams from two coronal
holes, a polar coronal hole above the CME and a smaller coronal hole
below. We have also interpreted evidence of compression at the back of
the CME at 1~AU as interaction from another coronal hole that rotated
beneath the CME. The location of the coronal hole and timing of the
interaction was consistent solar and heliospheric imaging and the
MESSENGER and STB in-situ observations. This analysis shows once more
the potential of combined multiview imaging and multipoint in-situ
measurements in deciphering the initiation and evolution of CME in
interplanetary space, even in the case of ''stealth' CMEs.

\acknowledgments
Drs. Nieves-Chinchilla, Vourlidas, and Stenborg are supported by the NASA SECCHI contract to NRL. The work of G.S. was also partly funded by NASA contract NNX11AD40G. The STEREO/SECCHI data used for this study are prepared by an international consortium of NASA Goddard Space Flight Center (USA), Lockheed Martin Solar and Astrophysics Lab (USA), Naval Research Laboratory USA), Rutherford Appleton Laboratory (UK), University of Birmingham (UK), Max-Planck-Institut f¬ur Sonnensystemforschung (Germany), Institut d'Optique Th`eorique et Appliqu`ee (France), Institut dÕAstrophysique Spatiale (France), and Centre Spatiale de Li'ege (Belgium). Dr. L.K. Jian's work is funded by NASA's Science Directorate as part of the STEREO project, including the IMPACT investigation. 
  
We also acknowledge the use of Wind, ACE and MESSENGER data available on NASA GSFC/SPDF (Space Physics Data Facility), and thank the STEREO teams for their open data policy.



\begin{thebibliography}{52}
\providecommand{\natexlab}[1]{#1}
\expandafter\ifx\csname urlstyle\endcsname\relax
  \providecommand{\doi}[1]{doi:\discretionary{}{}{}#1}\else
  \providecommand{\doi}{doi:\discretionary{}{}{}\begingroup
  \urlstyle{rm}\Url}\fi

\bibitem[{\textit{{Acu{\~n}a} et~al.}(2008)\textit{{Acu{\~n}a}, {Curtis},
  {Scheifele}, {Russell}, {Schroeder}, {Szabo}, and
  {Luhmann}}}]{Acunaetal2008SSRv}
{Acu{\~n}a}, M.~H., D.~{Curtis}, J.~L. {Scheifele}, C.~T. {Russell},
  P.~{Schroeder}, A.~{Szabo}, and J.~G. {Luhmann}, {The STEREO/IMPACT Magnetic
  Field Experiment}, \textit{\ssr}, \textit{136}, 203--226,
  \doi{10.1007/s11214-007-9259-2}, 2008.

\bibitem[{\textit{{Anderson} et~al.}(2007)\textit{{Anderson}, {Acu{\~n}a},
  {Lohr}, {Scheifele}, {Raval}, {Korth}, and {Slavin}}}]{Anderson2007}
{Anderson}, B.~J., M.~H. {Acu{\~n}a}, D.~A. {Lohr}, J.~{Scheifele}, A.~{Raval},
  H.~{Korth}, and J.~A. {Slavin}, {The Magnetometer Instrument on MESSENGER},
  \textit{\ssr}, \textit{131}, 417--450, \doi{10.1007/s11214-007-9246-7}, 2007.

\bibitem[{\textit{Bothmer and Schwenn}(1998)}]{Bothmer1998}
Bothmer, V., and R.~Schwenn, {The structure and origin of magnetic clouds in
  the solar wind}, \textit{Annales Geophysicae}, \textit{16}(1), 1--24, 1998.

\bibitem[{\textit{{Brueckner} et~al.}(1995)}]{Brueckner1995}
{Brueckner}, G.~E., et~al., {The Large Angle Spectroscopic Coronagraph
  (LASCO)}, \textit{\solphys}, \textit{162}, 357--402,
  \doi{10.1007/BF00733434}, 1995.

\bibitem[{\textit{{Burlaga} et~al.}(1981)\textit{{Burlaga}, {Sittler},
  {Mariani}, and {Schwenn}}}]{Burlaga1981}
{Burlaga}, L., E.~{Sittler}, F.~{Mariani}, and R.~{Schwenn}, {Magnetic loop
  behind an interplanetary shock - Voyager, Helios, and IMP 8 observations},
  \textit{\jgr}, \textit{86}, 6673--6684, \doi{10.1029/JA086iA08p06673}, 1981.

\bibitem[{\textit{{Burlaga} et~al.}(1998)}]{Burlaga1998JGR}
{Burlaga}, L., et~al., {A magnetic cloud containing prominence material -
  January 1997}, \textit{\jgr}, \textit{103}, 277, \doi{10.1029/97JA02768},
  1998.

\bibitem[{\textit{{Byrne} et~al.}(2010)\textit{{Byrne}, {Maloney}, {McAteer},
  {Refojo}, and {Gallagher}}}]{Byrne2010}
{Byrne}, J.~P., S.~A. {Maloney}, R.~T.~J. {McAteer}, J.~M. {Refojo}, and P.~T.
  {Gallagher}, {Propagation of an Earth-directed coronal mass ejection in three
  dimensions}, \textit{Nature Communications}, \textit{1}, 74,
  \doi{10.1038/ncomms1077}, 2010.

\bibitem[{\textit{{Davies} et~al.}(2012)}]{Davies2012}
{Davies}, J.~A., et~al., {A Self-similar Expansion Model for Use in Solar Wind
  Transient Propagation Studies}, \textit{\apj}, \textit{750}, 23,
  \doi{10.1088/0004-637X/750/1/23}, 2012.

\bibitem[{\textit{{DeForest} et~al.}(2013)\textit{{DeForest}, {Howard}, and
  {McComas}}}]{Deforest2013ApJ}
{DeForest}, C.~E., T.~A. {Howard}, and D.~J. {McComas}, {Tracking Coronal
  Features from the Low Corona to Earth: A Quantitative Analysis of the 2008
  December 12 Coronal Mass Ejection}, \textit{\apj}, \textit{769}, 43,
  \doi{10.1088/0004-637X/769/1/43}, 2013.

\bibitem[{\textit{{Delaboudini{\`e}re} et~al.}(1995)}]{Boudine1995}
{Delaboudini{\`e}re}, J.-P., et~al., {EIT: Extreme-Ultraviolet Imaging
  Telescope for the SOHO Mission}, \textit{\solphys}, \textit{162}, 291--312,
  \doi{10.1007/BF00733432}, 1995.

\bibitem[{\textit{{Galvin} et~al.}(2008)}]{Galvin2008SSRv}
{Galvin}, A.~B., et~al., {The Plasma and Suprathermal Ion Composition (PLASTIC)
  Investigation on the STEREO Observatories}, \textit{\ssr}, \textit{136},
  437--486, \doi{10.1007/s11214-007-9296-x}, 2008.

\bibitem[{\textit{{Garrard} et~al.}(1998)\textit{{Garrard}, {Davis}, {Hammond},
  and {Sears}}}]{Garradetal1998SSRv}
{Garrard}, T.~L., A.~J. {Davis}, J.~S. {Hammond}, and S.~R. {Sears}, {The ACE
  Science Center}, \textit{\ssr}, \textit{86}, 649--663,
  \doi{10.1023/A:1005096317576}, 1998.

\bibitem[{\textit{{Gopalswamy} et~al.}(2001)\textit{{Gopalswamy}, {Yashiro},
  {Kaiser}, {Howard}, and {Bougeret}}}]{Gopalswamy2001ApJ}
{Gopalswamy}, N., S.~{Yashiro}, M.~L. {Kaiser}, R.~A. {Howard}, and J.-L.
  {Bougeret}, {Radio Signatures of Coronal Mass Ejection Interaction: Coronal
  Mass Ejection Cannibalism?}, \textit{\apjl}, \textit{548}, L91--L94,
  \doi{10.1086/318939}, 2001.

\bibitem[{\textit{{Gopalswamy} et~al.}(2004)\textit{{Gopalswamy}, {Yashiro},
  {Krucker}, {Stenborg}, and {Howard}}}]{Gopalswamy2004JGR}
{Gopalswamy}, N., S.~{Yashiro}, S.~{Krucker}, G.~{Stenborg}, and R.~A.
  {Howard}, {Intensity variation of large solar energetic particle events
  associated with coronal mass ejections}, \textit{Journal of Geophysical
  Research (Space Physics)}, \textit{109}, A12105, \doi{10.1029/2004JA010602},
  2004.

\bibitem[{\textit{{Gopalswamy} et~al.}(2005)\textit{{Gopalswamy}, {Yashiro},
  {Michalek}, {Xie}, {Lepping}, and {Howard}}}]{Goplaswamy2005GRL}
{Gopalswamy}, N., S.~{Yashiro}, G.~{Michalek}, H.~{Xie}, R.~P. {Lepping}, and
  R.~A. {Howard}, {Solar source of the largest geomagnetic storm of cycle 23},
  \textit{\grl}, \textit{32}, L12S09, \doi{10.1029/2004GL021639}, 2005.

\bibitem[{\textit{{Gopalswamy} et~al.}(2009)\textit{{Gopalswamy},
  {M{\"a}kel{\"a}}, {Xie}, {Akiyama}, and {Yashiro}}}]{Gopalswamyetal2009JGRA}
{Gopalswamy}, N., P.~{M{\"a}kel{\"a}}, H.~{Xie}, S.~{Akiyama}, and
  S.~{Yashiro}, {CME interactions with coronal holes and their interplanetary
  consequences}, \textit{Journal of Geophysical Research (Space Physics)},
  \textit{114}, A00A22, \doi{10.1029/2008JA013686}, 2009.

\bibitem[{\textit{{Hidalgo} et~al.}(2002)\textit{{Hidalgo},
  {Nieves-Chinchilla}, and {Cid}}}]{Hidalgo2002a}
{Hidalgo}, M.~A., T.~{Nieves-Chinchilla}, and C.~{Cid}, {Elliptical
  cross-section model for the magnetic topology of magnetic clouds},
  \textit{\grl}, \textit{29}(13), 1637, \doi{10.1029/2001GL013875}, 2002.

\bibitem[{\textit{{Howard} et~al.}(2008)}]{Howard2008}
{Howard}, R.~A., et~al., {Sun Earth Connection Coronal and Heliospheric
  Investigation (SECCHI)}, \textit{\ssr}, \textit{136}, 67--115,
  \doi{10.1007/s11214-008-9341-4}, 2008.

\bibitem[{\textit{{Jian} et~al.}(2006)\textit{{Jian}, {Russell}, {Luhmann}, and
  {Skoug}}}]{Jian2006}
{Jian}, L., C.~T. {Russell}, J.~G. {Luhmann}, and R.~M. {Skoug}, {Properties of
  Interplanetary Coronal Mass Ejections at One AU During 1995 2004},
  \textit{\solphys}, \textit{239}, 393--436, \doi{10.1007/s11207-006-0133-2},
  2006.

\bibitem[{\textit{{Jian} et~al.}(2011)\textit{{Jian}, {Russell}, and
  {Luhmann}}}]{Jian2011}
{Jian}, L.~K., C.~T. {Russell}, and J.~G. {Luhmann}, {Comparing Solar Minimum
  23/24 with Historical Solar Wind Records at 1 AU}, \textit{\solphys},
  \textit{274}, 321--344, \doi{10.1007/s11207-011-9737-2}, 2011.

\bibitem[{\textit{{Kaiser} et~al.}(2008)\textit{{Kaiser}, {Kucera}, {Davila},
  {St.~Cyr}, {Guhathakurta}, and {Christian}}}]{Kaiser2008}
{Kaiser}, M.~L., T.~A. {Kucera}, J.~M. {Davila}, O.~C. {St.~Cyr},
  M.~{Guhathakurta}, and E.~{Christian}, {The STEREO Mission: An Introduction},
  \textit{\ssr}, \textit{136}, 5--16, \doi{10.1007/s11214-007-9277-0}, 2008.

\bibitem[{\textit{{Kilpua} et~al.}(2011)\textit{{Kilpua}, {Lee}, {Luhmann}, and
  {Li}}}]{Kilpua2011}
{Kilpua}, E.~K.~J., C.~O. {Lee}, J.~G. {Luhmann}, and Y.~{Li}, {Interplanetary
  coronal mass ejections in the near-Earth solar wind during the minimum
  periods following solar cycles 22 and 23}, \textit{Annales Geophysicae},
  \textit{29}, 1455--1467, \doi{10.5194/angeo-29-1455-2011}, 2011.

\bibitem[{\textit{{Lepping} et~al.}(1997)}]{Lepping1997JGR}
{Lepping}, R.~P., et~al., {The Wind magnetic cloud and events of October 18-20,
  1995: Interplanetary properties and as triggers for geomagnetic activity},
  \textit{\jgr}, \textit{102}, 14,049--14,064, \doi{10.1029/97JA00272}, 1997.

\bibitem[{\textit{{Lugaz}}(2010)}]{Lugaz2010SoPh}
{Lugaz}, N., {Accuracy and Limitations of Fitting and Stereoscopic Methods to
  Determine the Direction of Coronal Mass Ejections from Heliospheric Imagers
  Observations}, \textit{\solphys}, \textit{267}, 411--429,
  \doi{10.1007/s11207-010-9654-9}, 2010.

\bibitem[{\textit{{Lugaz} et~al.}(2009)\textit{{Lugaz}, {Vourlidas}, and
  {Roussev}}}]{Lugaz2009}
{Lugaz}, N., A.~{Vourlidas}, and I.~I. {Roussev}, {Deriving the radial
  distances of wide coronal mass ejections from elongation measurements in the
  heliosphere - application to CME-CME interaction}, \textit{Annales
  Geophysicae}, \textit{27}, 3479--3488, \doi{10.5194/angeo-27-3479-2009},
  2009.

\bibitem[{\textit{{Lugaz} et~al.}(2012)\textit{{Lugaz}, {Farrugia}, {Davies},
  {M{\"o}stl}, {Davis}, {Roussev}, and {Temmer}}}]{Lugaz2012ApJ}
{Lugaz}, N., C.~J. {Farrugia}, J.~A. {Davies}, C.~{M{\"o}stl}, C.~J. {Davis},
  I.~I. {Roussev}, and M.~{Temmer}, {The Deflection of the Two Interacting
  Coronal Mass Ejections of 2010 May 23-24 as Revealed by Combined in Situ
  Measurements and Heliospheric Imaging}, \textit{\apj}, \textit{759}, 68,
  \doi{10.1088/0004-637X/759/1/68}, 2012.

\bibitem[{\textit{{Luhmann} et~al.}(2008)}]{Luhmannetal2008_SSRv}
{Luhmann}, J.~G., et~al., {STEREO IMPACT Investigation Goals, Measurements, and
  Data Products Overview}, \textit{\ssr}, \textit{136}, 117--184,
  \doi{10.1007/s11214-007-9170-x}, 2008.

\bibitem[{\textit{{M{\"a}kel{\"a}} et~al.}(2013)\textit{{M{\"a}kel{\"a}},
  {Gopalswamy}, {Xie}, {Mohamed}, {Akiyama}, and {Yashiro}}}]{Makelaetal2013}
{M{\"a}kel{\"a}}, P., N.~{Gopalswamy}, H.~{Xie}, A.~A. {Mohamed}, S.~{Akiyama},
  and S.~{Yashiro}, {Coronal Hole Influence on the Observed Structure of
  Interplanetary CMEs}, \textit{\solphys}, p.~8,
  \doi{10.1007/s11207-012-0211-6}, 2013.

\bibitem[{\textit{{McComas} et~al.}(1998)\textit{{McComas}, {Bame}, {Barker},
  {Feldman}, {Phillips}, {Riley}, and {Griffee}}}]{McComasetal1998SSRv}
{McComas}, D.~J., S.~J. {Bame}, P.~{Barker}, W.~C. {Feldman}, J.~L. {Phillips},
  P.~{Riley}, and J.~W. {Griffee}, {Solar Wind Electron Proton Alpha Monitor
  (SWEPAM) for the Advanced Composition Explorer}, \textit{\ssr}, \textit{86},
  563--612, \doi{10.1023/A:1005040232597}, 1998.

\bibitem[{\textit{{M{\"o}stl} et~al.}(2012)}]{Mostl2012ApJ}
{M{\"o}stl}, C., et~al., {Multi-point Shock and Flux Rope Analysis of Multiple
  Interplanetary Coronal Mass Ejections around 2010 August 1 in the Inner
  Heliosphere}, \textit{\apj}, \textit{758}, 10,
  \doi{10.1088/0004-637X/758/1/10}, 2012.

\bibitem[{\textit{{Nieves-Chinchilla} et~al.}(2012)\textit{{Nieves-Chinchilla},
  {Colaninno}, {Vourlidas}, {Szabo}, A., and {et al.}}}]{Nieves-Chinchilla2012}
{Nieves-Chinchilla}, T., R.~{Colaninno}, A.~{Vourlidas}, {Szabo}, R.~A.,
  {Lepping}, and {et al.}, {Remote and in situ observations of an unusual
  Earth-directed Coronal Mass Ejection from multiple viewpoints},
  \textit{\jgr}, \textit{(accepted)}, 2012.

\bibitem[{\textit{{Ogilvie} et~al.}(1995)}]{Ogilvie1995}
{Ogilvie}, K.~W., et~al., {SWE, A Comprehensive Plasma Instrument for the Wind
  Spacecraft}, \textit{\ssr}, \textit{71}, 55--77, \doi{10.1007/BF00751326},
  1995.

\bibitem[{\textit{{Osherovich} and {Burlaga}}(1997)}]{Osherovich1997}
{Osherovich}, V., and L.~F. {Burlaga}, {Magnetic clouds}, \textit{Washington DC
  American Geophysical Union Geophysical Monograph Series}, \textit{99},
  157--168, \doi{10.1029/GM099p0157}, 1997.

\bibitem[{\textit{{Robbrecht} et~al.}(2009)\textit{{Robbrecht}, {Patsourakos},
  and {Vourlidas}}}]{Robbrecht2009}
{Robbrecht}, E., S.~{Patsourakos}, and A.~{Vourlidas}, {No Trace Left Behind:
  STEREO Observation of a Coronal Mass Ejection Without Low Coronal
  Signatures}, \textit{\apj}, \textit{701}, 283--291,
  \doi{10.1088/0004-637X/701/1/283}, 2009.

\bibitem[{\textit{{Savani} et~al.}(2009)\textit{{Savani}, {Rouillard},
  {Davies}, {Owens}, {Forsyth}, {Davis}, and {Harrison}}}]{Savani2009}
{Savani}, N.~P., A.~P. {Rouillard}, J.~A. {Davies}, M.~J. {Owens}, R.~J.
  {Forsyth}, C.~J. {Davis}, and R.~A. {Harrison}, {The radial width of a
  Coronal Mass Ejection between 0.1 and 0.4 AU estimated from the Heliospheric
  Imager on STEREO}, \textit{Annales Geophysicae}, \textit{27}, 4349--4358,
  \doi{10.5194/angeo-27-4349-2009}, 2009.

\bibitem[{\textit{{Savani} et~al.}(2010)\textit{{Savani}, {Owens}, {Rouillard},
  {Forsyth}, and {Davies}}}]{Savani2010}
{Savani}, N.~P., M.~J. {Owens}, A.~P. {Rouillard}, R.~J. {Forsyth}, and J.~A.
  {Davies}, {Observational Evidence of a Coronal Mass Ejection Distortion
  Directly Attributable to a Structured Solar Wind}, \textit{\apjl},
  \textit{714}, L128--L132, \doi{10.1088/2041-8205/714/1/L128}, 2010.

\bibitem[{\textit{{Savani} et~al.}(2011)\textit{{Savani}, {Owens}, {Rouillard},
  {Forsyth}, {Kusano}, {Shiota}, and {Kataoka}}}]{Savani2011}
{Savani}, N.~P., M.~J. {Owens}, A.~P. {Rouillard}, R.~J. {Forsyth},
  K.~{Kusano}, D.~{Shiota}, and R.~{Kataoka}, {Evolution of Coronal Mass
  Ejection Morphology with Increasing Heliocentric Distance. I. Geometrical
  Analysis}, \textit{\apj}, \textit{731}, 109,
  \doi{10.1088/0004-637X/731/2/109}, 2011.

\bibitem[{\textit{{Savani} et~al.}(2012)\textit{{Savani}, {Shiota}, {Kusano},
  {Vourlidas}, and {Lugaz}}}]{Savani2012}
{Savani}, N.~P., D.~{Shiota}, K.~{Kusano}, A.~{Vourlidas}, and N.~{Lugaz}, {A
  Study of the Heliocentric Dependence of Shock Standoff Distance and Geometry
  using 2.5D Magnetohydrodynamic Simulations of Coronal Mass Ejection Driven
  Shocks}, \textit{\apj}, \textit{759}, 103, \doi{10.1088/0004-637X/759/2/103},
  2012.

\bibitem[{\textit{{Sheeley} et~al.}(1999)\textit{{Sheeley}, {Walters}, {Wang},
  and {Howard}}}]{sheeley1999JGR}
{Sheeley}, N.~R., J.~H. {Walters}, Y.-M. {Wang}, and R.~A. {Howard},
  {Continuous tracking of coronal outflows: Two kinds of coronal mass
  ejections}, \textit{\jgr}, \textit{104}, 24,739--24,768,
  \doi{10.1029/1999JA900308}, 1999.

\bibitem[{\textit{{Sheeley} et~al.}(2008)}]{Sheeley2008ApJ}
{Sheeley}, N.~R., Jr., et~al., {Heliospheric Images of the Solar Wind at
  Earth}, \textit{\apj}, \textit{675}, 853--862, \doi{10.1086/526422}, 2008.

\bibitem[{\textit{{Smith} et~al.}(1998)\textit{{Smith}, {L'Heureux}, {Ness},
  {Acu{\~n}a}, {Burlaga}, and {Scheifele}}}]{Smithetal1998SSRv}
{Smith}, C.~W., J.~{L'Heureux}, N.~F. {Ness}, M.~H. {Acu{\~n}a}, L.~F.
  {Burlaga}, and J.~{Scheifele}, {The ACE Magnetic Fields Experiment},
  \textit{\ssr}, \textit{86}, 613--632, \doi{10.1023/A:1005092216668}, 1998.

\bibitem[{\textit{{Solomon} et~al.}(2001)}]{Solomon2001}
{Solomon}, S.~C., et~al., {The MESSENGER mission to Mercury: scientific
  objectives and implementation}, \textit{\planss}, \textit{49}, 1445--1465,
  \doi{10.1016/S0032-0633(01)00085-X}, 2001.

\bibitem[{\textit{{Stenborg} et~al.}(2008)\textit{{Stenborg}, {Vourlidas}, and
  {Howard}}}]{Stenborg2008}
{Stenborg}, G., A.~{Vourlidas}, and R.~A. {Howard}, {A Fresh View of the
  Extreme-Ultraviolet Corona from the Application of a New Image-Processing
  Technique}, \textit{\apj}, \textit{674}, 1201--1206, \doi{10.1086/525556},
  2008.

\bibitem[{\textit{{Szabo}}(1994)}]{Szabo1994}
{Szabo}, A., {An improved solution to the 'Rankine-Hugoniot' problem},
  \textit{\jgr}, \textit{99}, 14,737, \doi{10.1029/94JA00782}, 1994.

\bibitem[{\textit{{Thernisien} et~al.}(2006)\textit{{Thernisien}, {Howard}, and
  {Vourlidas}}}]{Thernisien2006}
{Thernisien}, A.~F.~R., R.~A. {Howard}, and A.~{Vourlidas}, {Modeling of Flux
  Rope Coronal Mass Ejections}, \textit{\apj}, \textit{652}, 763--773,
  \doi{10.1086/508254}, 2006.

\bibitem[{\textit{{Vinas} and {Scudder}}(1986)}]{Vinas1986}
{Vinas}, A.~F., and J.~D. {Scudder}, {Fast and optimal solution to the
  'Rankine-Hugoniot problem'}, \textit{\jgr}, \textit{91}, 39--58,
  \doi{10.1029/JA091iA01p00039}, 1986.

\bibitem[{\textit{{Vourlidas} et~al.}(2011)\textit{{Vourlidas}, {Colaninno},
  {Nieves-Chinchilla}, and {Stenborg}}}]{Vourlidas2011}
{Vourlidas}, A., R.~{Colaninno}, T.~{Nieves-Chinchilla}, and G.~{Stenborg},
  {The First Observation of a Rapidly Rotating Coronal Mass Ejection in the
  Middle Corona}, \textit{\apjl}, \textit{733}, L23,
  \doi{10.1088/2041-8205/733/2/L23}, 2011.

\bibitem[{\textit{{Wood} et~al.}(2010)\textit{{Wood}, {Howard}, and
  {Socker}}}]{Wood2010ApJ}
{Wood}, B.~E., R.~A. {Howard}, and D.~G. {Socker}, {Reconstructing the
  Morphology of an Evolving Coronal Mass Ejection}, \textit{\apj},
  \textit{715}, 1524--1532, \doi{10.1088/0004-637X/715/2/1524}, 2010.

\bibitem[{\textit{{Wood} et~al.}(2012)\textit{{Wood}, {Wu}, {Rouillard},
  {Howard}, and {Socker}}}]{Wood2012ApJ}
{Wood}, B.~E., C.-C. {Wu}, A.~P. {Rouillard}, R.~A. {Howard}, and D.~G.
  {Socker}, {A Coronal Hole's Effects on Coronal Mass Ejection Shock Morphology
  in the Inner Heliosphere}, \textit{\apj}, \textit{755}, 43,
  \doi{10.1088/0004-637X/755/1/43}, 2012.

\bibitem[{\textit{{Xie} et~al.}(2006)\textit{{Xie}, {Gopalswamy}, {Ofman},
  {St.~Cyr}, {Michalek}, {Lara}, and {Yashiro}}}]{Xie2006SpWea}
{Xie}, H., N.~{Gopalswamy}, L.~{Ofman}, O.~C. {St.~Cyr}, G.~{Michalek},
  A.~{Lara}, and S.~{Yashiro}, {Improved input to the empirical coronal mass
  ejection (CME) driven shock arrival model from CME cone models},
  \textit{Space Weather}, \textit{4}, S10002, \doi{10.1029/2006SW000227}, 2006.

\bibitem[{\textit{{Xie} et~al.}(2009)\textit{{Xie}, {St.~Cyr}, {Gopalswamy},
  {Yashiro}, {Krall}, {Kramar}, and {Davila}}}]{Xie2009}
{Xie}, H., O.~C. {St.~Cyr}, N.~{Gopalswamy}, S.~{Yashiro}, J.~{Krall},
  M.~{Kramar}, and J.~{Davila}, {On the Origin, 3D Structure and Dynamic
  Evolution of CMEs Near Solar Minimum}, \textit{\solphys}, \textit{259},
  143--161, \doi{10.1007/s11207-009-9422-x}, 2009.

\bibitem[{\textit{{Xie} et~al.}(2013)\textit{{Xie}, {Gopalswamy}, and
  {St.~Cyr}}}]{Xie2013}
{Xie}, H., N.~{Gopalswamy}, and O.~C. {St.~Cyr}, {Near-Sun Flux-Rope Structure
  of CMEs}, \textit{\solphys}, p.~6, \doi{10.1007/s11207-012-0209-0}, 2013.

\end{thebibliography}

\clearpage

\clearpage





\begin{table}
\begin{center}
\caption{Summary of the solar transient events (STE) as observed by the in-situ set of instruments onboard on MESSENGER, ACE and STEREO B spacecraft during the dates 2009 August 27--31.}
\label{Table:In-situ}
\begin{tabular}{cccccccccc}
\tableline \tableline
s/p  & STE$_{start}$ & FR$_{start}$ & STE$_{end}$ & B$_{max}$ & $<$\textbf{B}$>$ & $<$V$_{sw}>$ & V$_{exp}$ \\
 &  & doy MM/DD HH:MM  &  & nT & nT & km/s & km/s \\
\tableline
 MES& 239 08/27 17:02 &240 08/28 02:23&240 08/28 21:36 &21.8 &15.6 &  &   \\
 Earth & 242 08/30 00:16 &242 08/30 08:40&242 08/30 14:24 &12.7 &11.9 &401  &7.5   \\
 STB & 242 08/30 02:50 &242 08/30 16:20&243 08/31 08:24 &12.2 &8.6 &328  &16.6 & \\
 \tableline
\end{tabular}
\end{center}
\end{table}

\clearpage

\clearpage

\begin{table}
\begin{center}
\caption{Elements position on the Sun disk on Stonyhurst coordinates system.}
\label{Table:SunElem}
\begin{tabular}{llllll}
\tableline\tableline
Time&Element & Position& Long & Lat \\
\tableline

Aug 25$^{th}$ 02:13 &FP1& & 279 &42&  \\
& CH1& Central Point & 276 & 32  &  \\
& CH2& Central Point & 317 & 35  &  \\
\tableline
Aug 25$^{th}$ 12:14 &FP2 & & 14 &3 & \\
& CH1& Central Point & 283 & 31  &  \\
& CH2& Central Point & 322 & 32  &  \\
\tableline
\end{tabular}
\end{center}
\end{table}

\clearpage

\begin{table}
\begin{center}
\caption{In-situ analysis of the ICME elements as observed by each spacecraft. a) Shock direction, starting with the spacecraft, time, normal vector (n), longitude ($\phi_{shock}$) and latitude ($\theta_{shock}$) of normal vector. b) Flux-rope axis orientation;  longitude ($\phi_{FR}$), latitude ($\theta_{FR}$) and propagation angle ($\xi_{FR}$). The cross-section deformation (CS$_{def}$) and the impact parameter ($y_{0}$ related to the major axis $R_{max}$). Longitude is defined in the Sun-Earth plane with the zero pointed to the Sun, and latitude is counter-clockwise positive.}%
\label{Table:orientation}


\begin{tabular}{ccccccc}
a)  &  &  &  &  &  &\\
\tableline\tableline
s/p  &  & Time (HH:MM)  & $\overrightarrow{n}_{RTN}$ & $\phi_{shock}$ & $\theta_{shock}$ &\\
\tableline
 STB & shock$_{front}$  & 242 08/30 02:50 &  (0.997,0.077,-0.028) & $184^{\circ}$ & $\sim 0^{\circ}$ &   \\
 STB & shock$_{inside}$ & 242 08/30 03:41 & (0.960,-0.278,0.018) & $164^{\circ}$ & $\sim 0^{\circ}$ &   \\
 Wind & shock$_{front}$ & 242 08/30 00:16 &  (0.851,0.415,-0.321) & $206^{\circ}$ & $-19^{\circ}$ &  \\
 \tableline
 \\
 \newline
\end{tabular}
\begin{tabular}{cccccccc}
b) &  &  &  &  &  &  & \\
\tableline\tableline
s/p  &  Time (HH:MM) & Dur & $\phi_{FR}$ & $\theta_{FR}$ & $\xi_{FR}$ & CS$_{def}$ & y$_{0}$/R$_{max}$ \\
\tableline
MESS & 08/28 04:45 & $\sim 11$ hr & $251^{\circ}$ & $10^{\circ}$  & $42^{\circ}$ & 0.40 & 0.22 \\
STB  & 08/30 19:41 & $\sim 13$ hr & $253^{\circ}$ & $3^{\circ}$   & $44^{\circ}$ & 0.23 & 0.80 \\
ACE  & 08/30 08:40 & $\sim 6$ hr  & $309^{\circ}$ & $-15^{\circ}$ & $69^{\circ}$ & 0.30 & 0.08 \\
 \tableline
\end{tabular}

\end{center}

\end{table}

\clearpage







\end{document}